\documentclass[11pt, a4paper]{article}

\usepackage{color}
\usepackage{graphicx}
\usepackage{amssymb}
\usepackage{amsmath}
\usepackage[english]{babel}
\usepackage[T1]{fontenc}
\usepackage[noadjust]{cite}
\usepackage{hyperref} 

\numberwithin{equation}{section}

\usepackage{booktabs}
\usepackage{mdframed}
\usepackage{xcolor}

\usepackage{bbold}
\usepackage[utf8]{inputenc}
\usepackage{authblk}

\newcommand\eeq{\end{equation}}
\newcommand\beq{\begin{equation}}
\newcommand\eea{\end{eqnarray}}
\newcommand\bea{\begin{eqnarray}}

\begin{document}

\linespread{1.1}

\title{{\bf  The Standard Model partial unification scale\\ as a guide to new physics model building}}

\author[1,2]{ {\large\sc Isabella Masina} \thanks{masina@fe.infn.it}}
\author[3]{ {\large\sc Mariano Quir\'os} \thanks{quiros@ifae.es}}

\affil[1]{\small\it Dept.\,of Physics and Earth Science, Ferrara University, Via Saragat 1, 44122 Ferrara, Italy }
\affil[2]{\small\it Istituto Nazionale di Fisica Nucleare (INFN), Sez.\,di Ferrara, Via Saragat 1, 44122 Ferrara, Italy }
\affil[3]{\small\it Institut de F\'{\i}sica d'Altes Energies (IFAE) and The Barcelona Institute of  Science and Technology (BIST), Campus UAB, 08193 Bellaterra (Barcelona) Spain}

\date{}

\maketitle
\vskip .8cm

\begin{abstract}
In the Standard Model, partial unification of the non-Abelian running gauge couplings is achieved at the scale $\mu^{\rm SM}_{32} \approx 2.8 \times 10^{16}$ GeV.
Elaborating on this fact, we discuss a simple general parametrization  for the new physics corrections leading to full unification at some scale $M_X$.
We show that for any new physics model such that the corrections to the non-Abelian couplings are equal (or nearly so), 
$M_X$ is equal (or close to) the partial unification scale $\mu^{\rm SM}_{32}$; the latter scales could be disentangled only if the corrections to the non-Abelian couplings are significantly different. 
We explore how the parametrization  works for some relevant models with new physics below $M_X$, as low energy supersymmetry, split supersymmetry, \textit{etc}.
As for models with a desert up to $M_X$, we explore in particular how the parametrization works for string inspired corrections; 
we find a phenomenologically remarkable possibility for unification at about $100$ TeV, suggesting a low string scale, 
in addition to the more conservative possibility for unification at $\mu^{\rm SM}_{32}$; for models with power-low running/threshold corrections, we also outline an interesting connection with the number of fermion families propagating in the bulk.
\end{abstract}

\linespread{1.4}

\vskip 1.2cm
\newpage
\section{Introduction}

It is well known that gauge coupling unification (GCU) is missed in the Standard Model (SM), 
while it is achieved in the case of its low energy supersymmetric (SUSY) completion~\cite{Dimopoulos:1981yj, Dimopoulos:1981zb, Ibanez:1981yh, Sakai:1981gr, Einhorn:1981sx, Marciano:1981un}: 
the minimal supersymmetric Standard Model (MSSM) with superpartners at the TeV scale; 
this has indeed been one of the major hints in favor of low energy SUSY\,\cite{Amaldi:1991cn, Ellis:1990wk, Langacker:1991an, Giunti:1991ta}. 
For recent reviews about GCU in low energy SUSY (or other) models, we refer the interested reader to the PDG review on {\it {Grand Unified Theories}}\,\cite{ParticleDataGroup:2024cfk}\,
and to Ref.\,\cite{Senjanovic:2023jvv}.

There is an interesting coincidence that we want to point out at the beginning of this work, and on which we are going to further elaborate.
In the SM, the partial unification of the non-Abelian running gauge couplings, $\alpha_3(\mu)$ and $\alpha_2(\mu)$, 
is achieved at the scale $\mu^{\rm SM}_{32} \approx 2.8 \times 10^{16}$\,GeV, 
which is interestingly close to the (nearly full) GCU scale for the MSSM, $M^{\rm SUSY}_X \sim 2 \times 10^{16}$\,GeV.  
As will be discussed, the closeness of the two scales\,\footnote{At difference, in the context of string theory, Ref.~\cite{Dienes:1996du} discussed various attempts to identify the SM partial unification and string scales.} can actually be traced back to the values of the beta functions.
A similar situation also happens for the split-SUSY \cite{Arkani-Hamed:2004ymt,Giudice:2004tc} and for the 2-Higgs-Doublet model (2HDM) \cite{Lee:1973iz, Branco:2011iw} scenarios:
their non-Abelian coupling partial unification scales are close to the SM one.

We will use a simple general parametrization  for the new physics corrections leading to full (or non-Abelian partial) GCU at the scale $M_X$, 
which is suitable for any model beyond the SM (BSM).
It is based on the introduction of three parameters $\epsilon_i$ ($i=1,2,3$), encoding the deviations from the corresponding SM gauge couplings evaluated at $M_X$.
Depending on the relative magnitude of the non-Abelian corrections, $\epsilon_2$ and $\epsilon_3$, two kinds of possible scenarios arise:

\textit{i)} If the corrections to the non-Abelian gauge couplings are equal (or nearly so), the GCU scale $M_X$ turns out to be equal (or close to) 
the SM partial unification scale $\mu^{\rm SM}_{32}$. 
This is precisely what happens in the case of low energy SUSY, and is the reason behind the closeness of $M_{X}^{\rm SUSY}$ and $\mu^{\rm SM}_{32}$.
However, as will be shown, this also happens for the non-Abelian partial unification in other relevant models, including split-SUSY and 2HDM.
Hence, BSM models such that $\epsilon_2 \simeq \epsilon_3$ can be denoted as {\it mirage} SUSY,  
as they predict $M_X \simeq \mu^{\rm SM}_{32} \approx M_X^{\rm SUSY}$, as in the low energy SUSY case.

\textit{ii)} At the contrary, only if the corrections to the non-Abelian gauge couplings are significantly different, it is possible to disentangle $M_X$ from $\mu^{\rm SM}_{32}$; 
also in this case, it is possible to draw some interesting general considerations about the magnitude of $M_X$.

The proposed parametrization  is useful to study any BSM model providing GCU, without or with a desert up to $M_X$: 

- For models \textit{without a desert}, GCU is in general achieved because 
new physics below $M_X$ is able to suitably modify the beta functions, as happens for instance in low energy SUSY, 
split-SUSY, but also in many non-SUSY models; the new physics below or around $M_X$ also introduces (in general subdominant) threshold corrections to GCU, 
accounting for the decoupling of BSM particles. 
For a representative (necessarily incomplete) selection of works along these directions see {\it e.g.}~Refs.~\cite{Langacker:1992rq, Carena:1993ag, Altarelli:2000fu, Masina:2001pp, Kehagias:2005vz, DiLuzio:2013dda, Ellis:2015jwa, Schwichtenberg:2018cka, Meloni:2019jcf, Djouadi:2022gws, Haba:2024lox}.

- For models \textit{with a desert}, GCU relies on corrections at $M_X$ which are instead of ultra-violet (UV) origin, in the sense that the new physics is above $M_X$. 
Examples of UV origin corrections include: 
\textit{a)} string inspired corrections~\cite{Dienes:1996du, Cho:1997gm}, and;
\textit{b)} effective corrections induced by an additional non-renormalizable kinetic term in the Lagrangian~\cite{Hill:1983xh, Shafi:1983gz, Panagiotakopoulos:1984wf, Hall:1992kq}.

In this work, we first briefly show how to exploit the proposed parametrization  in the non-desert case.
The latter has been extensively analyzed in the literature by means of a variety of models; here we focus on a selection of the most representative models. 
Secondly, we analyze in detail how the parametrization  works for the desert case. For case \textit{a)},
we find interesting possibilities for GCU, both for $M_X=\mu^{\rm SM}_{32}$ and for $M_X \simeq 100$ TeV.  
In the class of models with power-low running~\cite{Dienes:1998vh,Dienes:1998vg,Dienes:1998qh}, we outline an interesting connection with the number of fermion families propagating in the bulk.
The desert case \textit{b)} requires a longer discussion, which will be addressed elsewhere.
 
The paper is organized as follows.
In Sec.\,\ref{sec-partial} we discuss the origin of the coincidence between the SM partial unification scale and the low energy SUSY unification scale, as well as other non-desert models, while in Sec.~\ref{sec-full} we analyze the requirements for full unification within the different models.
In Sec.\,\ref{sec-general} we introduce the general parametrization  for BSM corrections leading to GCU.
In Sec.\,\ref{sec-beta} we apply it to the non-desert case, where GCU is achieved by modifying the beta functions, 
often also considering threshold effects from decouplings of BSM particles.
In Sec.\,\ref{sec-stringy} we consider the desert case with corrections of stringy origin.
We finally draw our conclusions in Sec.\,\ref{sec-concl}.

\section{Partial unification in the SM and beyond}
\label{sec-partial}

As it is well known (see for instance the PDG review on {\it Grand Unified Teories} ~\cite{ParticleDataGroup:2024cfk} and references therein), 
the gauge couplings run at one-loop according to
\beq
\frac{1}{\alpha_i(\mu)}=\frac{1}{\alpha_i(\mu_0)} - \frac{b_i}{2 \pi} \, \log \left(  \frac{\mu}{\mu_0} \right) \,\, ,
\label{eq-running}
\eeq 
where  $\mu$ is the renormalization scale, $\mu_0$ is some low energy scale for matching with the experimentally derived values of the gauge couplings, 
and $b_i$ (with $i=1,2,3$) are the corresponding beta functions coefficients. Here we use the $SU(5)$ normalization for the hypercharge, that is $\alpha_1=5/3\, \alpha_Y$.

The general expression for the one-loop beta function of a gauge theory based on the symmetry group $G_i$ is
\beq
b_i=-\frac{11}{3}C_2(G_i)+\left(\frac{2}{3}n_f +\frac{1}{3}n_s \right)C(r)  \,,
\label{eq-betaSM}
\eeq
where $n_f$ is the number of chiral fermions and $n_s$ the number of complex scalars in the representation $r$ of $G_i$; 
for $SU(N)$, the quadratic Casimir is $C_2(SU(N))=N$ and $C(N)=1/2$ for the fundamental representation; 
for $U(1)$, $C_2(U(1))=0$ and $C(r)=Y^2$, where $Y$ is the hypercharge.
In particular, for a supersymmetric theory 
\beq
b_i^{\rm SUSY}=-3C_2(G_i)+n_r C(r) \,,
\label{eq-betaMSSM}
\eeq
where $n_r$ is the number of chiral multiplets in the representation $r$ of $G_i$, including the fermion-sfermion and Higgs-Higgsino sectors.

In the rest of this section, we will compare partial unification in the SM with partial unification in other distinguished non-desert models with extra stuff below $M_X$. 

\subsection{The SM vs.~the MSSM}

In the MSSM the region between the electroweak and unification scales is populated by the supersymmetric partners and the extra Higgs doublet, 
required to achieve anomaly cancellation induced by the Higgsinos.
By using the $SU(5)$ normalization for the hypercharge, the beta functions for the SM and the MSSM are respectively
\beq
b^{\rm SM}_i =(41/10,-19/6,-7) \,\, , \,\,\, b^{\rm MSSM}_i =(33/5,1,-3) \,\, .
\label{eq-beta-SMvsMSSM}
\eeq

\begin{figure}[htb!]
\vskip .5cm 
 \begin{center}
 \includegraphics[width=11.5 cm]{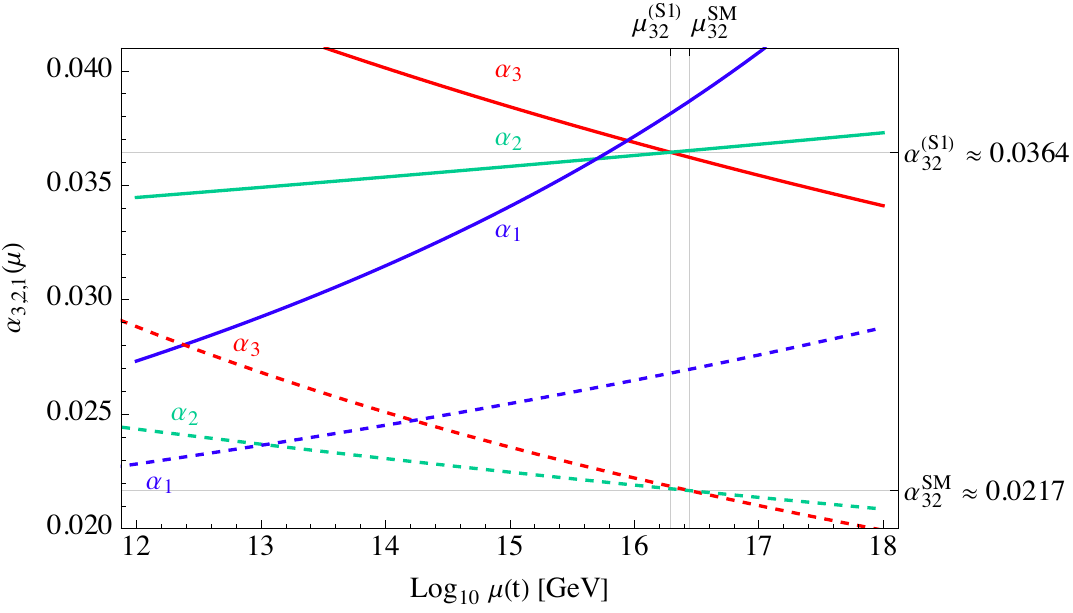}  \,\,\,\,
 \end{center}
\caption{\baselineskip=12 pt \small \it
Running gauge couplings in the \textrm{SM} at NNLO (dashed) and in the \textrm{MSSM} at LO (solid). The SUSY running starts at $10\, {\rm TeV}$. } 
\label{fig-SM}
\vskip .5 cm
\end{figure}

Using a NNLO calculation\,\footnote{It corresponds to at least three loops for the beta functions and at least two loops for the matching conditions.
Here, as done in \cite{Masina:2024ybn} to which we refer the reader for more details, we exploit the matching conditions at $200$\,GeV provided in \cite{Alam:2022cdv}.}, 
the running SM gauge couplings can be determined with high precision. 
For the strong coupling with five flavors, we consider the range $\alpha_3(m_Z)=0.1179 \pm 0.0009$, and fix the top quark and Higgs masses to their central values,
$m_t=172.5$ GeV and $m_H=125.25$ GeV \cite{ParticleDataGroup:2024cfk}; the experimental errors on the top quark and Higgs masses are indeed negligible for the sake of the present analysis. 
It turns out that the partial unification scale for the non-Abelian couplings and the corresponding unified coupling strength are
\beq
\mu^{\rm SM}_{32} = (2.8 \pm 0.3) \times 10^{16}\, {\rm GeV}\,, \,\,\, \alpha^{\rm SM}_{32} \equiv\alpha_{2}(\mu^{\rm SM}_{32})= \alpha_{3}(\mu^{\rm SM}_{32}) = 0.02166 \pm 0.00002 \,,
\eeq
where the error quoted corresponds to the 1$\sigma$ variation in $\alpha_3(m_Z)$.
The SM running couplings obtained by using the above mentioned central values as input parameters are represented by the dashed lines in Fig.~\ref{fig-SM}.

The LO (one loop) approximation for the running is also of interest, since it allows for a quick comparison among models (as will be done in the following). 
In the SM, taking the central values for the experimental inputs,
the partial unification scale at LO turns out to be $\mu^{(1)}_{32} \simeq 8.8  \times 10^{16}$\,GeV, as shown in the upper left panel of Fig.\,\ref{fig-ia} (dashed lines); 
in this case, the theoretical error in the calculation 
dominates over the one associated to the 1$\sigma$ variation in $\alpha_3(m_Z)$.

As for low energy SUSY, we simplify the calculation by fixing a common mass threshold for all supersymmetric particles, $\mu_S$, 
and some uncertainty is expected in the result depending on its value. 
Starting the SUSY running at $\mu_S=1(10)$\,TeV, the LO calculation gives $\mu^{(S1)}_{32}/10^{16}\, {\rm GeV} \simeq 2.05 (1.95)$ for the scale where the non-Abelian couplings meet;
the Abelian coupling coming close (the closer the smaller $\mu_S$), this suggests that full unification might be achieved by including subdominant effects. 
The common value of the non-Abelian gauge couplings turns out to be significantly larger than in the SM, $\alpha_{32}^{(S1)}\approx 0.0385 (0.0364)$.
The case $\mu_S=10$ TeV is explicitly shown in Fig.~\ref{fig-SM} (solid lines). 
Clearly, the unification scale would slightly change by considering two-loop beta functions, or including SUSY threshold corrections. 
We do not consider them, as this is an unnecessary refinement for the sake of the discussion in the present work.
Independently of the sophistication in the calculation, what is indeed interesting is that the partial unification scale for the non-Abelian gauge couplings in the SM and in the MSSM are close to each other. 

Is this fact just a coincidence, or something deeper is hidden?
Let us argue about this, starting by explicitly writing the condition for partial unification at $M_X$, namely $\alpha^{-1}_2(M_X)=\alpha^{-1}_3(M_X)$.
Using Eq.\,(\ref{eq-running}) with $\mu \rightarrow M_X$ and $\mu_0 \rightarrow \mu_S$,
the condition gives
\beq
\frac{1}{\alpha_2(\mu_S)} - \frac{1}{\alpha_3(\mu_S)} = \frac{b_2-b_3}{2 \pi} \log \left(  \frac{M_X}{\mu_S} \right) \,.
\label{eq-b2b3}
\eeq
Once $\mu_S$ is fixed, the left-hand side above is the same for the SM and the MSSM; the logarithmic term in the right-hand side is nearly equal, 
due to the previously discussed closeness of $\mu^{(1)}_{32}$ and $\mu^{(S1)}_{32}$. 
What might in principle be different is the quantity $b_2-b_3$, but 
we already know that there is going to be a mild difference.
Indeed, from Eq.\,(\ref{eq-beta-SMvsMSSM}), one has for the SM and the MSSM 
\beq
b^{\rm SM}_2-b_3^{\rm SM} = -\frac{19}{6} - (-7) = \frac{23}{6} \lesssim 4 \,, \,\,\,\, 
b_2^{\rm MSSM}-b_3^{\rm MSSM} = 1- (-3) = 4  \,.
\label{eq-par-MSSM}
\eeq
The fractional difference in $b_2-b_3$ for the two models is just $(1/6)/4=1/24$, which explains the closeness of the partial unification scales at one-loop,
$\mu^{(1)}_{32}$ and $\mu^{(S1)}_{32}$.

But where does the little difference come from?
From Eqs.\,(\ref{eq-betaSM}) and (\ref{eq-betaMSSM}), we see that
\beq
b_i^{\rm MSSM} -  b_i^{\rm SM}= \frac{2 }{3} C_2(G_i)  + \frac{1}{6} {\Delta n_f^{(i)}} + \frac{1}{6} \left( \Delta n_{\tilde f}^{(i)} + \Delta n_{h}^{(i)} \right) \, \,,
\eeq
where $\Delta n_f^{(i)}$, $\Delta n_{\tilde f}^{(i)}$ and $\Delta n_{h}^{(i)}$ are respectively the number of additional chiral fermions, real scalar partners of fermions (or sfermions) and Higgses which are present in the MSSM, on top of the SM, and belong to the fundamental representation of $G_i$.
Since $\Delta n_f^{(2)}=4$ for the Higgsinos, $\Delta n_{\tilde f}^{(2)}=9+3=12$ for the doublets of squarks and sleptons, and $\Delta n_{h}^{(2)}=1$ for the additional Higgs doublet, 
one has  
\beq
b_2^{\rm MSSM} -  b_2^{\rm SM}= \frac{2 }{3} \cdot 2 +\frac{1}{6} \underbrace{\Delta n_f^{(2)}}_{=4 } +\frac{1}{6} \underbrace{\Delta n_{\tilde f}^{(2)}}_{=12}  +\frac{1}{6} \underbrace{\Delta n_h^{(2)}}_{=1}= 4 +1/6 \, ,
\label{eq-MSSM-2}
\eeq
while, since $\Delta n_f^{(3)}=\Delta n_h^{(3)}=0$ and $\Delta n_{\tilde f}^{(3)}=12$ for the color triplets of squarks, one has
\beq
b_3^{\rm MSSM} -  b_3^{\rm SM}= \frac{2 }{3} \cdot 3+\frac{1}{6} \underbrace{ \Delta n_{\tilde f}^{(3)}}_{=12}   = 4 \,.
\label{eq-MSSM-3}
\eeq
 As a result, the sfermion contribution disappears in the difference below,
\beq
(b_2^{\rm MSSM}-b_3^{\rm MSSM}) - (b^{\rm SM}_2-b_3^{\rm SM}) =  \underbrace{\textrm{ gauginos} }_{-2/3} + \underbrace{ \textrm{Higgsinos} }_{2/3} + \underbrace{ \textrm{ add.~Higgs} }_{1/6} = 1/6 \,,
\label{eq-betadiff}
\eeq
where there is also perfect cancellation between the $-2/3$ from the gauginos and the $2/3$ from the Higgsinos (provided they have the same mass), 
but there is no way to cancel the $1/6$ from the additional scalar Higgs. 

The previously mentioned difference in the quantity $b_2-b_3$ (see the right-hand side of Eq.\,(\ref{eq-b2b3}))
can thus be traced back in the MSSM to the contribution of the additional Higgs scalar to $b_2$. 
Therefore, if we imagine to decouple the second MSSM Higgs scalar (not the corresponding Higgsino which is required for anomaly cancellation), 
the difference $b_2-b_3$ at one-loop would become equal for both the MSSM and the SM, so that at LO in both models $\alpha_2$ and $\alpha_3$ would meet at the same scale, 
$M_X=\mu_{32}^{(1)}$.

\subsection{Other models with partial unification close to $\mu^{\rm SM}_{32}$}

It is interesting to consider other models with extra stuff at $\mu_S$, between the electroweak scale and $M_X$, and whose partial unification scale turns out to be close to $\mu^{\rm SM}_{32}$.
This is the case for the 2HDM and for split-SUSY.

\subsubsection{The 2HDM}
The simplest model with extra matter below the unification scale is the 2HDM, with an extra Higgs doublet, also known as
non-supersymmetric 2HDM, whose beta functions are such that
\beq
b_2^{\rm 2HDM}=-3, \quad b_3^{\rm 2HDM}=-7 \quad \Rightarrow\quad b_2^{\rm 2HDM}-b_3^{\rm 2HDM} = 4\, .
\eeq
The difference above is precisely the same as in the MSSM, see Eq.~(\ref{eq-par-MSSM}).
As a consequence, in both theories $\alpha_2$ and $\alpha_3$ meet at LO at the same scale, provided the mass of the extra Higgs and of the SUSY particles are also the same.   
For instance, taking $\mu_S=10$ TeV, the partial unification scale in the 2HDM is $M_X=\mu_{32}^{(S1)}$, as shown in the upper right panel of Fig.\,\ref{fig-ia}.
What changes between the MSSM and the 2HDM is of course the prediction for the non-Abelian gauge coupling strength at $M_X$, which in the 2HDM is very close to the SM one.
Another difference is that full unification is much worse in the 2HDM than in the MSSM.

\subsubsection{Split-SUSY}
\begin{figure}[thb!]
\vskip .5cm 
 \begin{center}
 \includegraphics[width=7.6 cm]{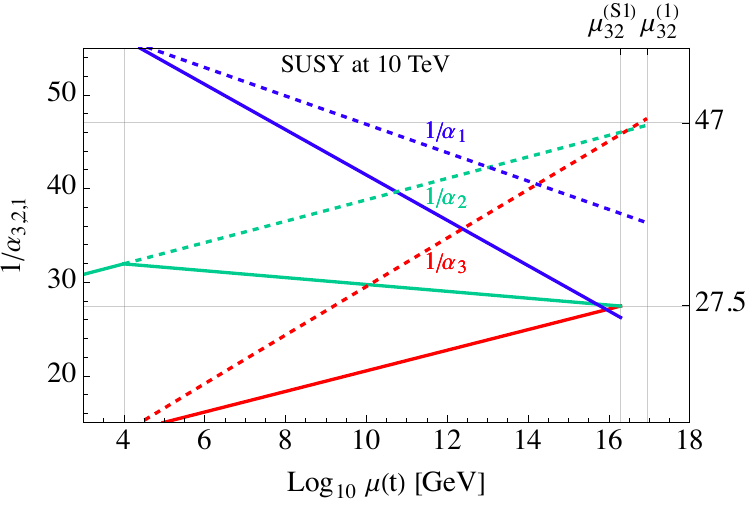}  \,\,\,\, \includegraphics[width=7.6 cm]{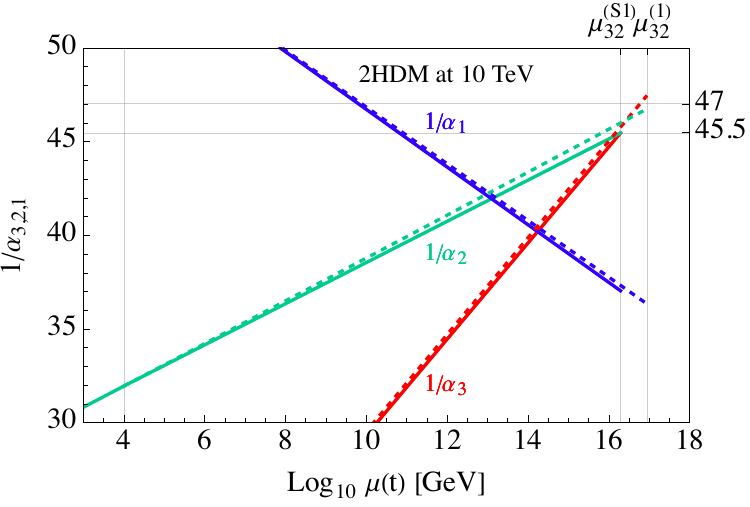}  \vskip .7cm  \includegraphics[width=7.6 cm]{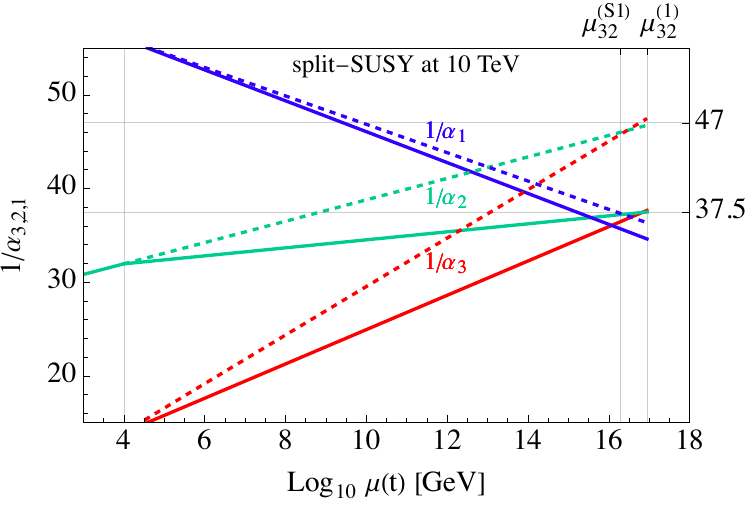}  
 \end{center}
\caption{\baselineskip=12 pt \small \it
Running gauge couplings at LO (solid) in the MSSM, 2HDM and split-SUSY, in the upper left, upper right and lower panel, respectively, with respect to the SM ones (dashed). 
The new physics running starts at the mass threshold  $\mu_S=10\, {\rm TeV}$. } 
\label{fig-ia}
\vskip 1.2 cm
\end{figure}

Another outstanding possibility, midway between the SM and the MSSM, is split-SUSY~\cite{Arkani-Hamed:2004ymt,Giudice:2004tc}, a theory where all scalars, except for the SM Higgs, 
are decoupled at the unification scale. In that case the theory below $M_X$ is the SM plus light gauginos and Higgsinos, 
an interesting possibility as the lightest supersymmetric particle can remain as a dark matter candidate in the presence of R-parity. 
In this theory the beta functions are such that
\beq
b_2^{\rm split-SUSY}=-\frac{7}{6},\quad b_3^{\rm split-SUSY}=-5\quad \Rightarrow\quad b_2^{\rm split-SUSY}-b_3^{\rm split-SUSY}=\frac{23}{6}\,.
\eeq
The difference above thus has the same value as 
in the pure SM. Indeed, inspecting Eq.\,(\ref{eq-betadiff}) we see that: the contribution from the sfermions disappears, so their mass scale is irrelevant
; the contributions from gauginos and Higgsinos cancel each other out, provided they are degenerate; and the $1/6$ from the additional Higgs is not present if its mass is larger than $M_X$.
If these conditions are fulfilled, at LO, the partial unification scale for split-SUSY is the same as in the SM, $M_X=\mu^{(1)}_{32}$.
This can be seen in the lower panel of Fig.\,\ref{fig-ia}, which shows the evolution of the gauge couplings for split-SUSY with $\mu_S=10$ TeV.
Notice also that full unification is a bit worse than in the MSSM.

\section{Full unification}
\label{sec-full}

It is straightforward to estimate the value of $b_1-b_3$ which would ensure full unification at $M_X$. 
Readapting Eq.\,(\ref{eq-b2b3}), we have, at LO, 
\beq
b_1 -b_3 = \left( \frac{1}{\alpha_1(\mu_S)} - \frac{1}{\alpha_3(\mu_S)} \right) \, \frac{2\pi}{ \log \left(  \frac{M_X}{\mu_{S}} \right)}  \, ,
\label{eq-b1b3}
\eeq
where $\mu_S$ is a threshold scale. 

We first focus on the pure SM: taking $M_X=\mu^{(1)}_{32}$, from Eq.~(\ref{eq-b1b3}) we obtain $b_1 -b_3\approx 8.9$.
Then, let us now focus on a BSM model where extra stuff is introduced at $\mu_S$,
always assuming that partial unification of the non-Abelian gauge couplings occurs at $M_X$, and 
taking for definiteness $\mu_S=10$\,TeV. From Eq.\,(\ref{eq-b1b3}) with $M_X=\mu^{(1)}_{32}$, one obtains $b_1 -b_3\approx 8.9$ (as in the SM); 
taking instead $M_X=\mu^{(S1)}_{32}$, one gets $b_1 -b_3\approx 9.3$ (as in the MSSM).

We can now compare the values obtained from Eq.\,(\ref{eq-b1b3}) with the corresponding values derived from the numerical values of beta functions in the SM and in the MSSM, 
Eq.\,(\ref{eq-beta-SMvsMSSM});
the agreement of the two values can be used as an indicator of the model success in providing full GCU. 
For example:\\
\noindent - in the SM, the beta functions give $b_1 -b_3 = 41/10 -(-7)= 111/10 =11.1$, 
which is sizably larger than the value $8.9$ required to achieve full unification at $M_X=\mu^{(1)}_{32}$; \\
\noindent - in the MSSM, the beta functions give $b_1 -b_3 = 33/5 -(-3) =9.6$, 
which is actually close to the value $9.3$ associated to full unification at $M_X=\mu^{(S1)}_{32}$. 

Let us also analyze the other relevant models already introduced in Sec.~\ref{sec-partial}: \\
\noindent - for the 2HDM, the Abelian beta function at LO is $b_1= 42/10$, so that $b_1 -b_3 = 42/10 -(-7)= 11.2$, 
to be compared with the value $9.3$ corresponding to full unification at $M_X=\mu^{(S1)}_{32}$. As a result, GCU does not work well in this model, anyway better than in the SM; \\
\noindent - for split-SUSY, the Abelian beta function at LO is $b_1=45/10$, so that $b_1 -b_3 = 45/10-(-5)= 9.5$, 
to be compared with $8.9$, associated to full unification at $M_X=\mu^{(1)}_{32}$. In this case GCU works quite well, but not as well as in the MSSM.

All these considerations are compatible with Fig. \ref{fig-ia}. 
Clearly, the previous results rely on adopting the $SU(5)$ normalization for the hypercharge.
We now briefly discuss the impact of having another normalization.

\subsubsection*{Hypercharge normalization}

As it is known, full GCU might be achieved even in the SM by using another hypercharge normalization, $\alpha'_1=k_1  \alpha_1$. 
By inspecting Fig.\,\ref{fig-SM}, in order to suitably suppress the SM (blue) dashed curve associated to the Abelian coupling, 
at NNLO one finds numerically that $k_1\approx 0.77$. 
Hence, in order to achieve full GCU in the SM, one would need $k_Y  \equiv  \alpha'_1/\alpha_Y = 5/3 \cdot k_1 \approx 1.28$.
The latter result is compatible with the one derived in~\cite{Dienes:1996du}, that is $k_Y =13/10$;
according to the discussion in\,\cite{Dienes:1996du}, such value is well within the constraints imposed by string theory.

Focussing on the SM and adopting the hypercharge normalization, $\alpha'_1= k_1 \alpha_1$, Eq.~(\ref{eq-b1b3}) is generalized as follows, 
\beq
b'_1 -b_3 =  \left( \frac{1}{k_1 \alpha_1(\mu_0)} - \frac{1}{\alpha_3(\mu_0)} \right) \, \frac{2\pi}{\log \left(  \frac{M_X}{\mu_{0}} \right)}  \approx 12.5\,,
\eeq
where the numerical value in the right-hand-side is obtained by taking $k_1 \approx 0.77$ and $M_X=\mu^{(1)}_{32}$.
As a consistency check, notice that using the SM beta functions provided in Eq.\,(\ref{eq-beta-SMvsMSSM}) and $k_1\approx 0.77$, one indeed has $b'_1-b_3= b_1/k_1-b_3 \approx 12.5$.

\section{General parametrization  for BSM corrections to GCU}
\label{sec-general}

Let us now encode the effect of any BSM model providing GCU at $\mu=M_X$ by writing 
\beq
\alpha_G\equiv  (1+\epsilon_1)\, \alpha^{\rm SM}_1(M_X) = (1+\epsilon_2)\, \alpha^{\rm SM}_2(M_X) = (1+\epsilon_3)\, \alpha^{\rm SM}_3(M_X) \,,
\label{eq-eps-gen}
\eeq
where the $\alpha^{\rm SM}$'s are the SM running couplings evolved at $M_X$ and the $\epsilon$'s encode the corrections due to the new physics able to provide GCU.
As already mentioned, the $\epsilon$'s could originate from many effects, as for instance: 
from the beta functions (as for low energy SUSY) and/or threshold corrections reflecting the mass spectrum of new heavy particles with masses slightly below $M_X$; 
from a $d=5$ gauge kinetic term; 
or from corrections due to relations of stringy origin.

In particular, for low energy SUSY (switched on at $10$ TeV as considered before) GCU occurs at $M_X=\mu^{(S1)}_{32}$, 
which is interestingly close to the SM partial unification scale $\mu^{\rm SM}_{32}$, as already discussed in Sec.\,\ref{sec-partial}; 
since by definition $ \alpha^{\rm SM}_2(\mu^{\rm SM}_{32})=\alpha^{\rm SM}_3(\mu^{\rm SM}_{32})$, 
Eq.\,(\ref{eq-eps-gen}) implies that $\epsilon^{\rm MSSM}_2 \approx \epsilon^{\rm MSSM}_3$
(we will come back to their small difference in the following).

Because of the same argument, 
any alternative model (for instance a dimension 5 operator modifying the gauge bosons kinetic term) such that $\epsilon^{\rm BSM}_2 \approx \epsilon^{\rm BSM}_3$, 
also provides unification at $M_X \approx \mu^{\rm SM}_{32}$. 
So, any model such that $\epsilon_2 \approx \epsilon_3$ can be said to be of the ``mirage SUSY'' type, 
in the sense that it might lead us thinking that low energy SUSY is truly realized in nature, while actually this might not be the case.
There is however a potential difference between mirage SUSY and true SUSY models: the value of $\alpha_G$. 
While in the MSSM $\alpha_G \simeq 0.0364$, as shown in Fig.\,\ref{fig-SM}, in other models it has to be computed and might be different. 
Of course, a place where the difference in $\alpha_G$ might manifest itself is proton decay.

Summarizing, one can distinguish two main scenarios, according to the relative size of their non-Abelian corrections to the gauge couplings, $\epsilon_2$ and $\epsilon_3$.
Only in the case of nearly equal correction (or mirage SUSY type models), the GCU scale is close to the SM partial unification scale, $M_X \approx \mu^{\rm SM}_{32}$. 

On the other hand, in the case of unequal corrections to the non-Abelian gauge couplings, we can figure out the impact on $M_X$.
By inspecting Fig.\,\ref{fig-SM}, we can visually realize that when $\epsilon_i$ is positive (negative), the combination $(1+\epsilon_i) \alpha^{\rm SM}_i(\mu)$ 
can be obtained by shifting up (down) the corresponding SM dashed curve.
The case $\epsilon_3 > \epsilon_2$ corresponds to a vertical separation of the curves for $(1+\epsilon_3) \alpha^{\rm SM}_3(\mu)$  
and $(1+\epsilon_2) \alpha^{\rm SM}_2(\mu)$ which is larger than in the SM case;
so that partial unification for the non-Abelian couplings is achieved at a scale larger than $\mu^{\rm SM}_{32}\approx 2.8 \times 10^{16}$ GeV.
On the contrary, when $\epsilon_3 < \epsilon_2$, the vertical separation gets reduced with respect to the SM, so that partial unification of the non-Abelian couplings 
happens at a smaller scale than $\mu^{\rm SM}_{32}$. 
Depending on the model, a high value for $M_X$ might possibly be preferred to prolong proton lifetime avoiding conflict with experimental constraints.

In the following, let us first study in more detail GCU for mirage SUSY models, that is models such that $\epsilon_3=\epsilon_2$, 
in which case $M_X$ is necessarily identified with the scale of SM partial unification $\mu^{\rm SM}_{32}$;
we will discuss later on models such that $\epsilon_3 \neq \epsilon_2$, in which case it should be possible to disentangle $M_X$ from $\mu^{\rm SM}_{32}$.

\subsection{GCU at $\mu^{\rm SM}_{32}$ (or mirage SUSY)}

\begin{figure}[t!]
\vskip .cm 
 \begin{center}
 \includegraphics[width=7.7 cm]{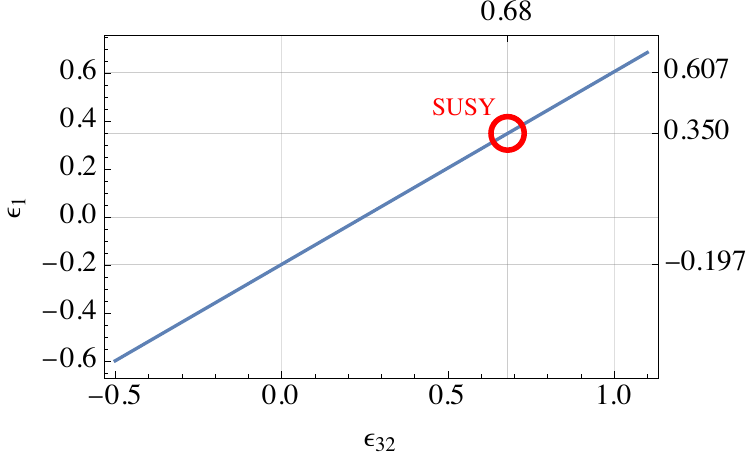}  \,\,\,\, \includegraphics[width=7.7 cm]{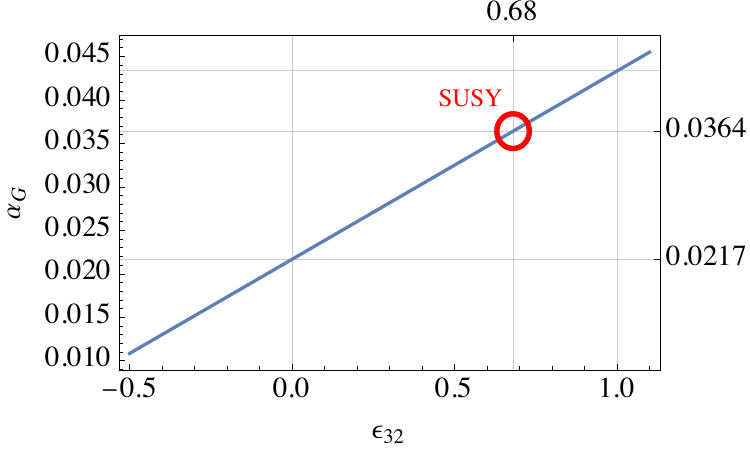}  \\
 \end{center}
\caption{\baselineskip=12 pt \small \it
The case of equal corrections to the non-Abelian gauge couplings. The dependence of $\epsilon_1$ (left) and $\alpha_G$ (right) with respect to $ \epsilon_{32}$.
Fits are provided in Eq.\,(\ref{eq-fit-e32}).} 
\label{fig-eps1eps32}
\vskip 1. cm
\end{figure}

As just discussed, GCU is achieved at $M_X=\mu^{\rm SM}_{32}$ if the corrections to the SM non-Abelian gauge couplings of Eq.\,(\ref{eq-eps-gen}) are equal
\beq
 \epsilon_{32} \equiv  \epsilon_3=\epsilon_2 \,.
\eeq 
From Eq.\,(\ref{eq-eps-gen}) one can derive how $\epsilon_1$ and $\alpha_G$ depend on $ \epsilon_{32}$, as shown in Fig.\,\ref{fig-eps1eps32}.
Actually the fits are extremely simple
\beq
\epsilon_1 \approx -0.197 + 0.804 \,\epsilon_{32} \,\, , \,\, \alpha_G \approx  0.0217\, (1+ \epsilon_{32})  \, .
\label{eq-fit-e32}
\eeq

\begin{figure}[h!]
\vskip 0.5cm 
 \begin{center}
 \includegraphics[width=7.3 cm]{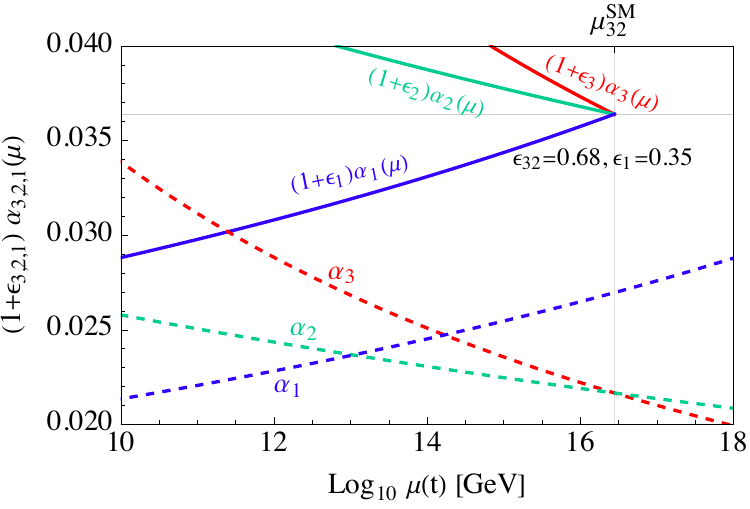} \,\,\, \,\,\,\includegraphics[width=7.3 cm]{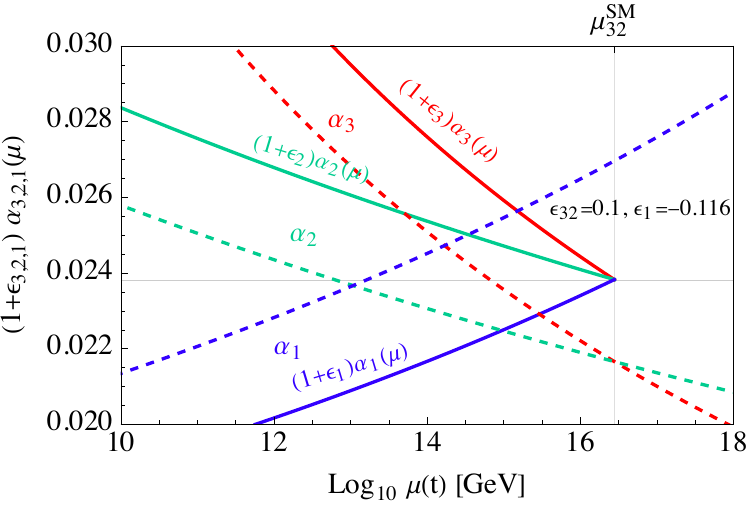}
 \end{center}
\caption{\baselineskip=12 pt \small \it
Examples of models with unification at $M_X=\mu^{\rm SM}_{32}$. Left panel: a model mimicking SUSY with $\epsilon_{32}=0.68$ and $\epsilon_1=0.35$, so that 
$\alpha_G = \alpha^{(S1)}_{32} \approx 0.0364$.
Right panel: a model of the mirage SUSY type, obtained taking $\epsilon_{32}=0.1$ and $\epsilon_1=-0.116$; in this case $\alpha_G =0.0238$.} 
\label{fig-ex-mirage}
\vskip 1. cm
\end{figure}

The fit of Eq.\,(\ref{eq-fit-e32}) allows to easily discuss some relevant cases.
For instance, notice that the SM with a different hypercharge normalization is recovered in the case $\epsilon_{32}=0$; in this case indeed 
$k_1=\alpha'_1/\alpha_1= 1+\epsilon_1 \approx 1 -0.197 = 0.803$.
The latter value is consistent with the result $k_1=0.77$, discussed in Sec.\,\ref{sec-full}.

Notice also that, since low energy SUSY is compatible with $\epsilon_2 \approx \epsilon_3$ and corresponds to $\alpha^{(S1)}_{32} \approx 0.0364$, 
we learn from Eq.\,(\ref{eq-fit-e32}) that it approximately corresponds to the values $\epsilon_{32} \approx 0.68$ and $\epsilon_1 \approx 0.35$.
In the left panel plot of Fig.\,\ref{fig-ex-mirage}, we show how SUSY would look like from the epsilon's point of view: the solid lines represent the combinations $(1+\epsilon_i) \alpha^{\rm SM}_i(\mu)$ (with $i=1,2,3$), which manifestly unify at $M_X=\mu^{\rm SM}_{32}$, as expected; the dashed lines represent the $\alpha^{\rm SM}_i(\mu)$; 
we omit the apex "SM" in the plots in order to simplify the notation. 

We now consider a specific example of GCU of the mirage SUSY type. It corresponds to taking $\epsilon_{32}=0.1$ and $\epsilon_1=-0.116$, 
and is shown in the right plot of Fig.\,\ref{fig-ex-mirage}.
While the dashed lines represent the $\alpha^{\rm SM}_i(\mu)$ (with $i=1,2,3$), solid lines represent the combinations $(1+\epsilon_i) \alpha^{\rm SM}_i(\mu)$, which manifestly unify at $M_X=\mu^{\rm SM}_{32}$, 
as expected. Notice that in this case, consistently with Eq.\,(\ref{eq-fit-e32}), $\alpha_{G}  \approx 0.0238$, 
significantly smaller than the value obtained with low energy SUSY (see Fig.\,\ref{fig-SM}).

Notice that also the 2HDM and split-SUSY are models of the mirage SUSY type, as their partial unification scale is equal or close to $\mu^{\rm SM}_{32}$. We will discuss the corresponding values of the epsilon's corrections, together with the value of the partially unified gauge couplings, $\alpha_{32}$, in the next section.

\subsection{GCU far from $\mu^{\rm SM}_{32}$}

In order to dissociate $M_X$ from $\mu^{\rm SM}_{32}$, one needs new physics capable of breaking the equality between the corrections to the non-Abelian gauge couplings, 
$\epsilon_2$ and $\epsilon_3$,
defined in Eq.\,(\ref{eq-eps-gen}). 
As already anticipated, $\epsilon_3 > \epsilon_2$ leads to $M_X > \mu^{\rm SM}_{32}$, while $\epsilon_3 < \epsilon_2$ to $M_X < \mu^{\rm SM}_{32}$. 

\begin{figure}[b!]
\vskip 1.5cm 
 \begin{center}
 \includegraphics[width=7.4 cm]{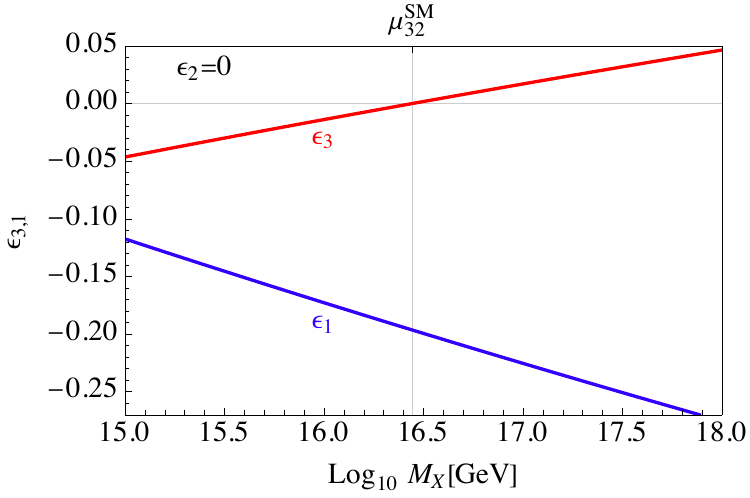}  \,\,\,\, \includegraphics[width=8.1 cm]{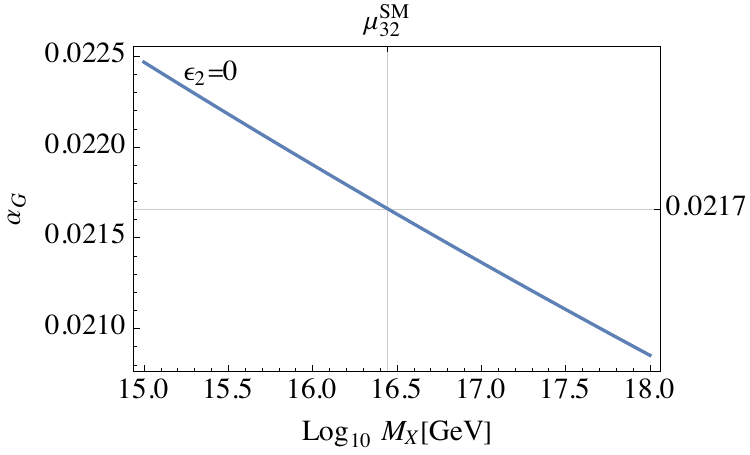}  
 \vskip 1.5cm
 \includegraphics[width=7.4 cm]{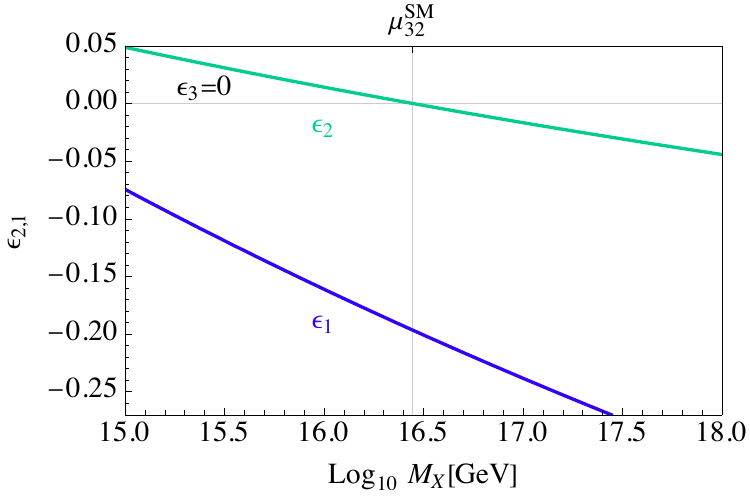}  \,\,\,\, \includegraphics[width=8.1 cm]{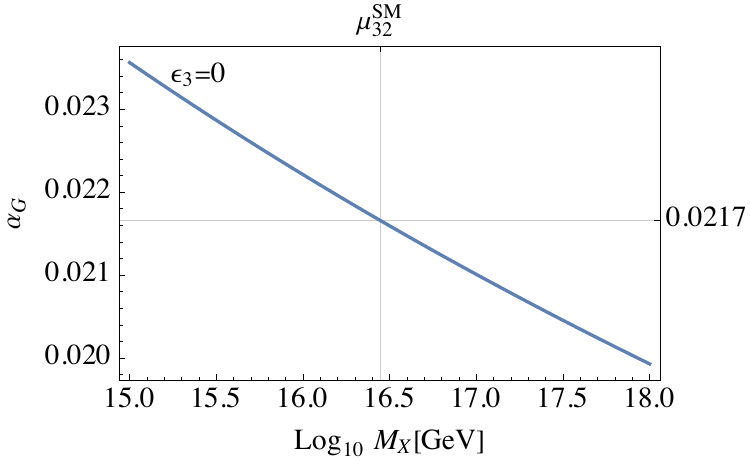}  
 \end{center}
\caption{\baselineskip=12 pt \small \it
The case of unequal corrections to the non-Abelian gauge couplings. Top panels: Case $\epsilon_2=0$: the dependence of $\epsilon_{3,1}$ and $\alpha_G$ on $M_X$.
Bottom panels: Case $\epsilon_3=0$: the dependence of $\epsilon_{2,1}$ and $\alpha_G$ on $M_X$. } 
\label{fig-eps230}
\vskip 1.5 cm
\end{figure}

Let us consider as an example the case $\epsilon_2=0$. 
We can calculate from Eq.\,(\ref{eq-eps-gen}) the values of $\epsilon_3$, $\epsilon_1$ and $\alpha_G$ which lead to GCU at the scale $M_X$, 
as shown in the top panels of Fig.\,\ref{fig-eps230}. 
As expected, we find that with positive (negative) and increasing $\epsilon_3$, $M_X$ increases (decreases) with respect to $\mu^{\rm SM}_{32}$.

Let us now turn to the case $\epsilon_3=0$. 
The values of $\epsilon_2$, $\epsilon_1$ and $\alpha_G$ which lead to GCU at the scale $M_X$ are shown in the bottom panels of Fig.\,\ref{fig-eps230}. 
Now, we find that with positive (negative) and increasing $\epsilon_2$, $M_X$ decreases (increases) with respect to $\mu^{\rm SM}_{32}$.

It is possible to write a more general formula for $M_X$ as a function of $\epsilon_3$ and $\epsilon_2$.
By explicitly writing the partial unification condition, $\alpha_{32} \equiv (1+\epsilon_2) \alpha^{\rm SM}_2(M_X) = (1+\epsilon_3) \alpha^{\rm SM}_3(M_X) $, 
and using Eq.\,(\ref{eq-running}) with $\mu_0 \rightarrow \mu^{\rm SM}_{32}$ and $\mu \rightarrow M_X$,
together with the definition $\alpha^{\rm SM}_2(\mu^{\rm SM}_{32})= \alpha^{\rm SM}_3(\mu^{\rm SM}_{32})=\alpha^{\rm SM}_{32}$, 
we obtain the dependence of $M_X$ on the non-Abelian epsilons
\beq
\frac{M_X} {\mu^{\rm SM}_{32}} = \, \exp \left( \frac{2 \pi}{\alpha^{\rm SM}_{32}}\, \frac{\epsilon_3 -\epsilon_2}{  (1+\epsilon_3)\, b^{\rm SM}_2 -(1+\epsilon_2)\, b^{\rm SM}_3  } \right) \,,
\label{eq-MX}
\eeq 
all the other parameters appearing in the right-hand-side being numerically well known. 
We can see the crucial exponential dependence on the difference $\epsilon_3-\epsilon_2$; if the latter vanishes we consistently recover the mirage SUSY result.  

\begin{figure}[h!]
\vskip 1. cm 
 \begin{center}
  \includegraphics[width=10. cm]{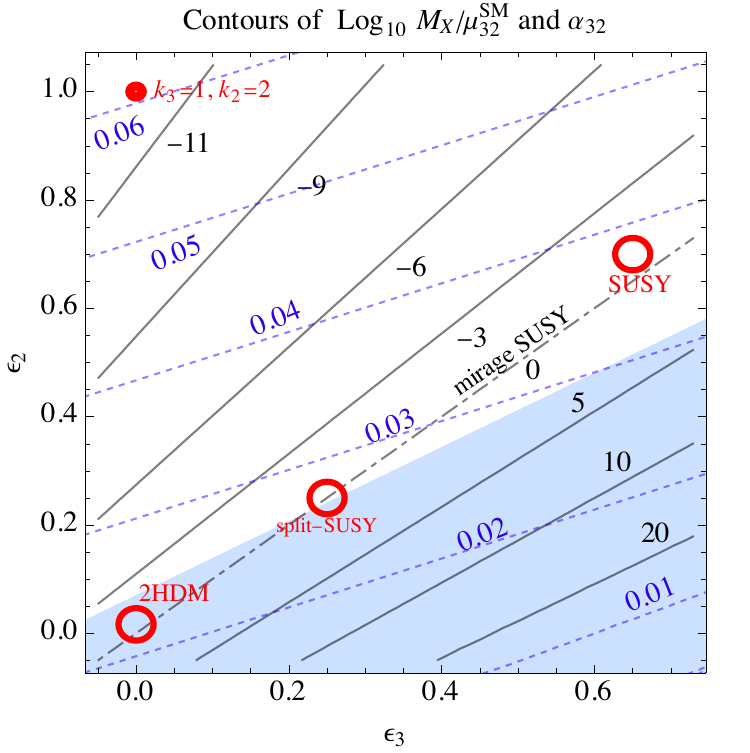}  
 \end{center}
\caption{\baselineskip=12 pt \small \it
Contours of ${\rm Log}_{10} M_X/\mu_{32}^{\rm SM}$ (solid black) and $\alpha_{32}$ (dashed blue). 
The circles (whose size is arbitrary) emphasize the values corresponding to selected models discussed in the text. The shaded (light blue) region is such that full unification would be achieved with $\epsilon_1 <0$.} 
\label{fig-MX}
\vskip 1.0 cm
\end{figure}

The general result is shown in the contour plot of Fig.\,\ref{fig-MX}, where also the contours of the partially unified couplings, $\alpha_{32}$, are displayed. 
For each point, there is a specific value of $\epsilon_1$ that would ensure full unification at $M_X$; the shaded (light blue) region corresponds to $\epsilon_1 <0$.
The circles emphasize the values corresponding to selected models, as discussed in the following. 
The size of the circles is arbitrary and serves just as a reminder of the uncertainty associated to our LO calculation, and to the subdominant effects associated to possible threshold corrections. 
Notice that the dot-dashed line in Fig.\,\ref{fig-MX} designates the values of the non-Abelian corrections corresponding to mirage SUSY models.

\section{GCU from beta functions}
\label{sec-beta}

In order to capture the main idea, let us consider in some details the simplified case in which there is new physics at very low energy, 
so that the beta functions are modified with respect to those of the SM at scales larger than the new physics scale, $\mu_S$.

Using Eq.~(\ref{eq-running}), we can write down the explicit dependence of the $\epsilon$'s in terms of the beta functions:
\beq
\epsilon_i =  b^{\rm BSM}_i  \frac{1 - \frac{1}{1+\epsilon_i} \frac{b^{\rm SM}_i}{b^{\rm BSM}_i} }{2\pi} \,  \alpha^{\rm SM}_i (M_X)\, \log \left( \frac{M_X}{\mu_S} \right) 
\approx \frac{b^{\rm BSM}_i  -b^{\rm SM}_i }{2\pi} \,  \alpha^{\rm SM}_i (M_X)\, \log \left( \frac{M_X}{\mu_S} \right) \,\,,
\label{eq-eps-beta}
\eeq
where the approximation in the right hand side is good provided $\epsilon_i \ll 1$ and/or $b^{\rm SM}_i/b^{\rm BSM}_i \ll 1+ \epsilon_i$.

Notice that, due to the Abelian nature of the hypercharge, Eq.~(\ref{eq-betaSM}) implies that $b_1^{\rm BSM}\geq b_1^{\rm SM}$; 
this means that in any BSM model with extra matter below the unification scale, $\epsilon_1\geq 0$. 
Then, from Fig.\,\ref{fig-MX} we see that it is not possible to achieve unification at the scale $\mu_{32}^{\rm SM}$ by just adding to the SM (which is located at $\epsilon_2=\epsilon_3=0$), 
extra matter (at scales below the unification scale) which is charged only under the hypercharge.

In the case in which the approximation of Eq.~(\ref{eq-eps-beta}) is valid, the condition $\epsilon_2 \approx \epsilon_3$ is thus
\beq
b^{\rm BSM}_3 - b^{\rm SM}_3 
\approx b^{\rm BSM}_2 - b^{\rm SM}_2 \,. 
\eeq
As an example, let us focus on low energy SUSY with $\mu_S =10$ TeV. 
Eqs.\,(\ref{eq-MSSM-2}) and (\ref{eq-MSSM-3}) lead to $b^{\rm MSSM}_i - b^{\rm SM}_i=4+\delta_{i2}/6$ for $i=2,3$ (where $\delta_{ij}$ is the Kronecker delta).
Using $M_X = \mu^{(S1)}_{32}=2 \times 10^{16}$ GeV, we have $\alpha_2(M_X) \approx \alpha_3(M_X)$.
From Eq.\,(\ref{eq-eps-beta}) we obtain that $\epsilon_2=0.70$ and $\epsilon_3=0.65$; these values are emphasized by means of a circle in Fig.\,\ref{fig-MX}. 
The latter values are very close, as expected, and justify to exploit the previous general discussion for $\epsilon_2=\epsilon_3$: 
in particular, from Fig.\,\ref{fig-eps1eps32} we see that for $\epsilon_{32}=0.68$, GCU is achieved with $\epsilon_1 \approx 0.35$ and, consistently, 
$\alpha_{G} \approx \alpha_{32}^{(S1)}=0.0364$.

As another example, let us consider the non-supersymmetric 2HDM where, as already discussed, the non-Abelian couplings meet at the same scale as for low energy SUSY, 
$M_X = \mu^{(S1)}_{32}$.
In this case 
\beq
b_2^{\rm 2HDM}-b_2^{\rm SM}= 1/6\,\, ,  \,\, b_3^{\rm 2HDM}-b_3^{\rm SM}=0\,.
\eeq
The beta functions of this model are very close to the SM ones; as the SM corresponds to $\epsilon_2=\epsilon_3=0$, here too we expect small values for the non-Abelian epsilons. 
Indeed, from Eq.\,(\ref{eq-eps-beta}) with $\mu_S=10$ TeV, we obtain $\epsilon_2=0.0167$ and $\epsilon_3=0$, as shown in Fig.\,\ref{fig-MX}; this model lies in the shaded (light blue) region,
where full unification would be achieved with $\epsilon_1 <0$.
We can exploit here the general discussion of the previous section:
having a look in particular to the bottom plots in Fig.\,\ref{fig-eps230}, we see that GCU requires $\epsilon_1 \approx -0.190$ and $\alpha_G \approx 0.022$.
As already commented in Sec.\,\ref{sec-general}, $b_1-b_3$ for the 2HDM is not compatible with full unification; the new physics required cannot just be charged under $U(1)_Y$, as in this case one would obtain $\epsilon_1>0$.

Similarly, for split-SUSY, we have
\beq
b_2^{\rm split-SUSY}-b_2^{\rm SM}= -7/6-(-19/6) =2  \,\, ,  \,\, b_3^{\rm split-SUSY}-b_3^{\rm SM}=-5-(-7)=2\,.
\eeq
Using $M_X = \mu^{(1)}_{32}$ with $\mu_S=10$ TeV, we have $\epsilon_2=\epsilon_3=0.25$, as shown by the corresponding circle in Fig.\,\ref{fig-MX}. 
GCU then requires $\epsilon_1 \approx 0.0012$ and $\alpha_G \approx 0.027$, as can be checked from Fig.\,\ref{fig-eps1eps32}.
We recall that, as shown in Sec.\,\ref{sec-general}, the value of $b_1-b_3$ for the split-SUSY case is compatible with full unification.

Clearly, it is possible to extend and refine the previous simplified analysis in various ways, to account for models where the desert is composed of various ``steps'' towards unification, and to include threshold corrections associated to the decoupling of BSM particles filling the desert. For a representative, although necessarily incomplete selection of models of this type, see for instance Refs.\,\cite{Langacker:1992rq, Carena:1993ag, Altarelli:2000fu, Masina:2001pp, Kehagias:2005vz, DiLuzio:2013dda, Ellis:2015jwa, Schwichtenberg:2018cka, Meloni:2019jcf, Djouadi:2022gws, Haba:2024lox}.

\section{String inspired unification}
\label{sec-stringy}

In string theories with non-standard Kac-Moody levels the gauge coupling unification happens at a scale $M_X$ such that
\beq
k_i \alpha_i(M_X)=\textrm{constant},\quad \textrm{where}\quad k_Y=\frac{5}{3}k_1 \,,
\eeq
where $k_i$ are the \textit{Kac-Moody} (or \textit{affine}) levels~\footnote{The affine levels have their origin in string theory. For instance in heterotic string constructions gauge symmetries $G_i$ are realized as affine world-sheet Lie algebras with central extensions realized at level $k_i$. For a review see e.g.~Ref.~\cite{Dienes:1996du}.}. While $k_2$ and $k_3$ are positive integers, the value of $k_1$ depends on the structure of the string theory and can in principle take any positive value.

Encoding the effect of any BSM model providing GCU with respect to the SM as in Eq.\,(\ref{eq-eps-gen}),
we can write the relation with Kac-Moody levels as
\beq
k_i \alpha_i(M_X)=(1+\epsilon_i) \alpha^{\rm SM}_i(M_X) \,.
\label{eq-KM-eps}
\eeq

Models with different values of $k_2$ and $k_3$ have been analyzed in Refs.~\cite{Dienes:1996du,Cho:1997gm}, 
although the case $k_2=k_3\equiv k$ is phenomenologically favored (as already known and will be numerically proven in the following). 

In the following two subsections, we first consider the case in which the $\alpha_i$ in Eq.\,(\ref{eq-KM-eps}) are identical to the SM ones, so that $k_i=1+\epsilon_i$;
in the last subsection, we provide a case in which the $\alpha_i$ are different from the SM ones as they include threshold effects.

\subsection{GCU at $\mu^{\rm SM}_{32}$ (or mirage SUSY)}

Identifying the $\alpha_i$ in Eq.\,(\ref{eq-KM-eps}) with those of the SM, we consider the case in which
\beq
\epsilon_{32} \equiv \epsilon_2=\epsilon_3=k-1\,,
\label{eq:k}
\eeq
where $k$ is a positive integer. Hence for $k=1$, $\epsilon_{32}=0$, while for $k=2$, $\epsilon_{32}=1$.
This is a string inspired realization of a mirage SUSY type model.

The value of $\epsilon_1$ is then given by $\epsilon_1=1-k_1$ and we have two free parameters: $k$ and $k_1$.
In all cases, it turns out that the unification scale is $M_X=\mu^{\rm SM}_{32}$ and
\beq
\alpha_{G}=k\, \alpha_2(\mu^{\rm SM}_{32})=k \, \alpha_3(\mu^{\rm SM}_{32})=k_1\, \alpha_1(\mu^{\rm SM}_{32}) \, ,
\eeq
from where the values of $\alpha_G$ and $k_1$ can be obtained, given the numerical value of $\alpha_{32}^{\rm SM}$ and fixing\,$k$.

\begin{figure}[htb!]
\vskip .5cm 
 \begin{center}
 \includegraphics[width=7.8 cm]{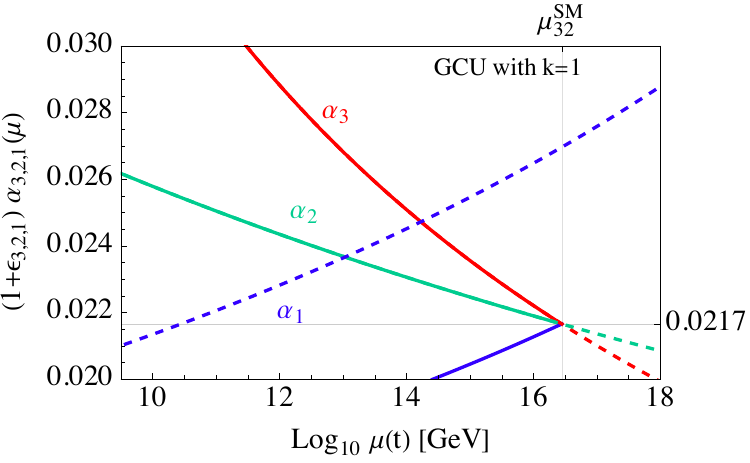}  \,\,\,\,
\includegraphics[width=7.5 cm]{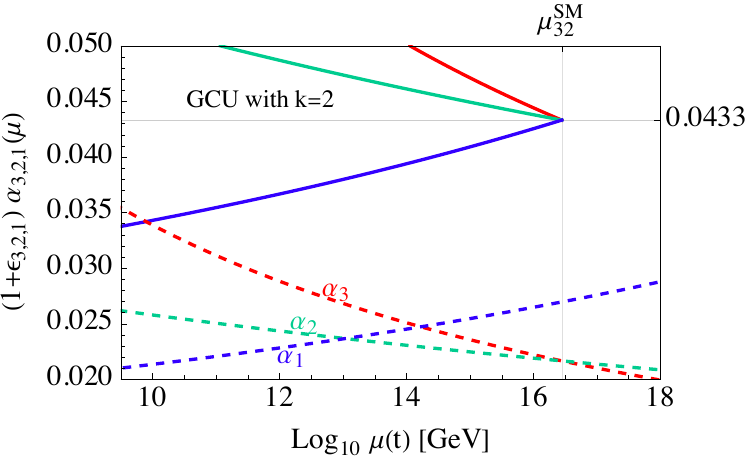}  
 \end{center}
\caption{\baselineskip=12 pt \small \it
Solid lines represent the quantities $(1+\epsilon_i) \alpha_i (\mu)$ in the case of string-inspired corrections leading to full GCU. 
Left panel: For $k=1$, the corrections are given by $\epsilon_2=\epsilon_3=0$, and $\epsilon_1=-0.197$. Right panel: For $k=2$, the corrections are $\epsilon_2=\epsilon_3=1$ and $\epsilon_1=0.607$.
The dashed lines stands for the SM running couplings, $\alpha_i (\mu)$.} 
\label{fig-k}
\vskip .5 cm
\end{figure}

In particular, for $k=1$, $\epsilon_2=\epsilon_3=0$, while $\epsilon_1=-0.197$. 
The associated quantities $(1+\epsilon_i) \alpha_i (\mu)$ are shown in the left plot Fig.\,\ref{fig-k};
notice that $\alpha_G$ is precisely the value of the SM partial unification, $\alpha_{32}^{\rm SM} \approx 0.0217$.
For $k=2$, $\epsilon_2=\epsilon_3=1$, while $\epsilon_1=0.607$. The associated quantities $(1+\epsilon_i) \alpha_i (\mu)$ are shown in the right plot of Fig.\,\ref{fig-k};
now $\alpha_G$ is definitely larger than $\alpha_{32}^{\rm SM}$.
As a check, notice these findings are also consistent with Fig.\,\ref{fig-eps1eps32}, taking the particular cases $\epsilon_{32}=0$ and $1$, respectively.

\subsection{GCU far from $\mu^{\rm SM}_{32}$} 

We again identify the $\alpha_i$ in Eq.\,(\ref{eq-KM-eps}) with those of the SM. In the case $k_2=1$ and $k_3=2$, we have $\epsilon_2=0$ and $\epsilon_3=1$. From the discussion in the previous section we thus expect $M_X >\mu^{\rm SM}_{32}$. 
Actually, from the upper left plot in Fig.\,\ref{fig-eps230} and from Fig.\,\ref{fig-MX}, we see that $M_X$ would be much larger than the Planck scale. This possibility is thus highly unfavored.

\begin{figure}[htb!]
\vskip .5cm 
 \begin{center}
  \includegraphics[width=7.3 cm]{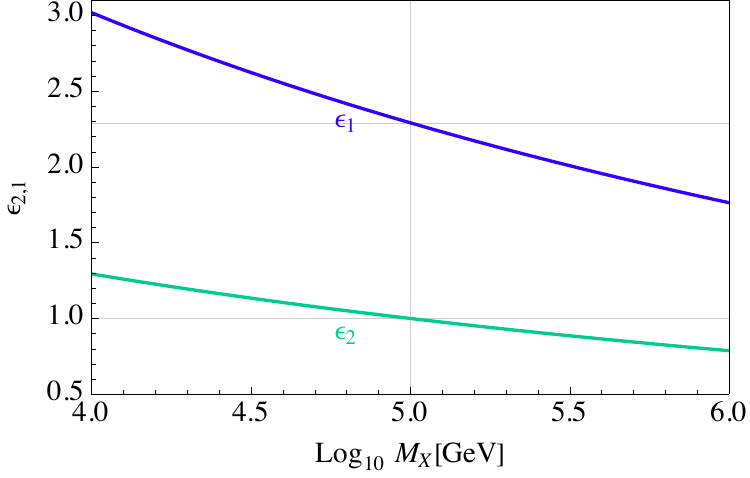}  \,\,\, 
 \includegraphics[width=8 cm]{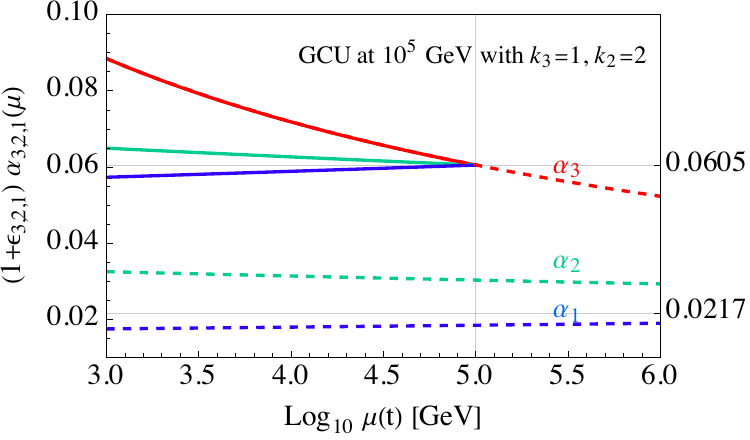}   
 \end{center}
\caption{\baselineskip=12 pt \small \it
String inspired corrections to GCU for $k_2=2$ and $k_3=1$, that is $\epsilon_2=1$ and $\epsilon_3=0$. 
Left: Zoom on the low energy region for $M_X$. Right: The quantities $(1+\epsilon_i) \alpha_i(\mu)$ unify at $M_X=10^5$ TeV when $\epsilon_1=2.29$.
The dashed lines stands for the SM running couplings, $\alpha_i (\mu)$.}
\label{fig-k2k3}
\vskip .5 cm
\end{figure}

At the contrary, in the case $k_2=2$ and $k_3=1$, we have $\epsilon_2=1$ and $\epsilon_3=0$. 
We thus expect $M_X < \mu^{\rm SM}_{32}$. Actually, from an extrapolation to low energy of the bottom left plot in Fig.\,\ref{fig-eps230}, we obtain the left plot of Fig.\,\ref{fig-k2k3}, 
where we see that, with $\epsilon_2=1$ and $\epsilon_1=2.29$, GCU happens at $M_X =10^5$\,GeV, as also expected from Fig.\,\ref{fig-MX}. 
The associated quantities $(1+\epsilon_i) \alpha_i (\mu)$ are represented by the solid lines in the right plot of Fig.\,\ref{fig-k2k3}, the dashed lines standing for the SM running couplings, $\alpha_i (\mu)$.
We are thus led to postulate the possibility of GCU at $100$\,TeV, with $\alpha_G \approx 0.0605$. 

In string theory there are essentially two ways of lowering the string scale $M_s$ from its natural value $\sim M_P$ (the Planck scale) to e.g.\,$\sim 100$ TeV, using the string relation with the 4D Planck scale
\beq
M_P^2=\frac{1}{g_s^2}M_s^8 V_6\,,
\eeq
where $g_s$ is the string scale and $V_6= (2 \pi R)^6$ the compactified volume, $R$ being the compactified radius.

For $g_s=\mathcal O(1)$ and $M_s\sim 100$ TeV, one can increase the compactification volume such that $1/R \sim 0.5$ GeV. This is the so-called ``large extra dimension scenario'' proposed in Refs.~\cite{Arkani-Hamed:1998jmv,Antoniadis:1998ig}. In this case the SM fields cannot propagate in the bulk of the extra dimensions, but be localized on 3-branes perpendicular to the transverse dimensions. Only the gravitational sector can propagate in the bulk making it difficult its experimental detection. Of course the size of the six extra dimensions can be different. In particular, the possibility of low string scale and two extra dimensions of a micron size has been recently considered in Refs.~\cite{Heckman:2024trz, Anchordoqui:2025nmb, Antoniadis:2025pet, Ettengruber:2025kzw, Antoniadis:2025rck}.

The other possibility is considering $M_s\sim 100$ TeV, while the inverse compactification radii are at the TeV scale. In this case the size of the 4D Planck scale is fixed by the smallness of the string coupling, $g_s\sim M_s^4 V_6^{1/2}/M_P\sim 10^{-10}$. The string construction which realizes this possibility is the so called Little String Theory, which is defined on a stack of NS-fivebranes where gravity is decoupled in the limit $g_s\to 0$ (for a review on Little String Theory see Ref.~\cite{Kutasov:2001uf}). In this case the SM fields can propagate in the bulk and as such their Kaluza-Klein excitations can be detected at present and future colliders. Moreover the string states with masses $\sim M_s$ could be searched for in future high-energy colliders, as FCC-hh~\cite{Helsens:2019bfw}. This possibility was first explored in Refs.~\cite{Antoniadis:2001sw,Antoniadis:2011qw} where all scales are considered at the TeV.

\subsection{Power-law unification}

Another possibility, 
motivated by a class of string theories (typically Type I or Type I$^{\prime}$ strings), is a $D>4$ dimensional theory with four flat dimensions and $\delta=D-4$ compactified dimensions, with a common compactification radius $R=M_c^{-1}$ much smaller than the unification scale, $M_c\ll M_X$, which is typically associated with the string scale $M_s$. For scales below $M_s$ the theory is an effective field theory defined in four flat and $\delta$ compactified dimensions. Concentrating on the gauge couplings, below $M_c$ they run logarithmically with the scale $\mu$ according to the 4D RGE. For scales $\mu>M_c$ the beta functions change every time a Kaluza-Klein (KK) threshold is crossed, and for a large number of thresholds the logarithmic running can be described by a power-law behavior ($\sim\mu^\delta$), which we can call, with an abuse of language, power-low ``running", but it is really a finite threshold effect at the corresponding scale $\mu$~\cite{Dienes:1998vh,Dienes:1998vg,Dienes:1998qh}. The large number of thresholds corresponding to KK states accelerate the behavior of gauge couplings which can then unify at much lower scales than the corresponding 4D theory.

Most applications of these theories deal with the case of $N=1$ supersymmetric theories at the level of zero modes, which implies $N=2$ supersymmetry at the level of KK modes~\cite{Dienes:1998vh,Dienes:1998vg,Delgado:1999ba}. The non-supersymmetric case~\cite{Dienes:1998vg} is much simpler as, at the level of KK modes, the contribution of gauge bosons is simply complemented by a scalar in the adjoint representation of the gauge group, while chiral fermions have extra partners with opposite chirality.

We will here concentrate, for simplicity, in the case of $\delta=1$ so that there is a 5D space compactified on the orbifold $S^1/\mathbb Z_2$, where all KK states have masses $m_n=m_0+n/R$ for $n\in Z$, and $m_0$ is the possible zero-mode mass, which is usually negligible as compared to the compactification masses. The $S^1/\mathbb Z_2$ orbifold has two fixed points at the two boundaries of the interval $[0,\pi R]$, where some 4D states can be localized. In general we will assume that there are propagating in the bulk the gauge bosons, the Higgs field and a number $\eta\leq 3$ of fermion generations. For the gauge fields, the four dimensional components $A_\mu$ are even under parity and the fifth component, the real scalar $A_5$, is odd (so no zero mode), both in the adjoint representation of the gauge group. Below $M_c$ only $A_\mu$ contributes to the gauge coupling running, while beyond $M_c$ the KK modes of both $A_\mu$ and $A_5$ do contribute to the running. As for fermions in the bulk, their KK modes are Dirac fermions while the zero modes are chiral, as one chirality is even and the opposite is odd, and thus without zero mode. Therefore, using the general expression (\ref{eq-betaSM}) the contribution of KK modes to the beta functions is given by
\beq
\tilde b_i=-\frac{7}{2}C_2(G_i)+\frac{1}{3}n_s C(r_s)+\frac{4}{3}n_f C(r_f)
\eeq
where $n_s$ ($n_f$) is the number of complex scalars (fermions) propagating in the bulk in the representation $r_s$ ($r_f$), and we have considered that KK fermions are Dirac. Also notice that KK modes from the real scalar $A_5$, in the adjoint representation, contributes as much as  $\frac{1}{6} C(Adj)=\frac{1}{6} C_2(G_i)$ to Eq.~(\ref{eq-betaSM}).
The running with the scale $\mu$ is then given, for $\delta=1$, by
\beq
\alpha_i^{-1}(\mu)=\alpha_i^{-1}(m_Z)-\frac{b_i}{2\pi}\log\frac{\mu}{m_Z}+
\left[\frac{\tilde b_i}{2\pi}\log\frac{\mu}{M_c}-\frac{\tilde b_i}{\pi} \left(\frac{\mu}{M_c}-1 \right) \right]\Theta(\mu-M_c)
\label{eq:alphas_DDG}
\eeq
where $\Theta(x)$ is the step function equal to 1 (0) for $x\geq 0$ ($x<0$), so as the KK modes only contribute for scales $\mu>M_c$.

For the SM with gauge bosons, Higgs and $\eta$ fermion generations propagating in the bulk, the zero modes ($b_i$) and  KK ($\tilde b_i$) beta functions are given by
\beq
(b_3,b_2,b_Y)=(-7,-19/6,41/6),\quad (\tilde b_3,\tilde b_2,\tilde b_Y)=\left(-\frac{21}{2},-\frac{41}{6},\frac{1}{6}\right)+\frac{8}{3}\eta\left(1,1,\frac{5}{3}\right)
\eeq 
and we are using as initial condition $\alpha_Y^{-1}(m_Z)=98.29$.

\begin{figure}[htb!]
\vskip .5cm 
 \begin{center}
  \includegraphics[width=7.7 cm]{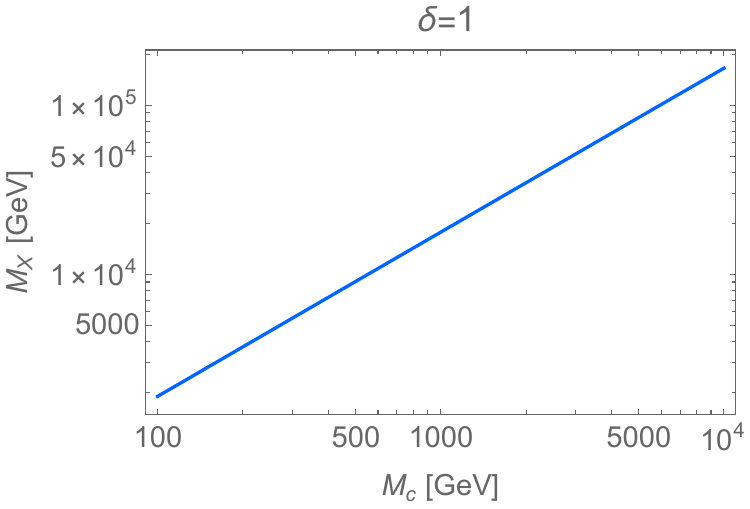}  \,\,\,\,\,
 \includegraphics[width=7.5 cm]{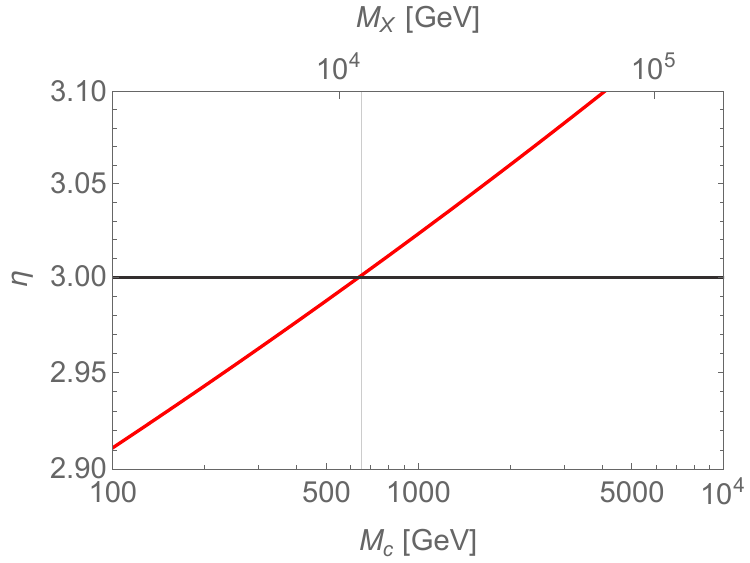}  
 \end{center}
\caption{\baselineskip=12 pt \small \it 
The partial unification scale $M_{X}$ (left panel) and the number of bulk generations $\eta$ ensuring full unification (right panel) as functions of the compactification scale $M_c$, for the case of one extra dimension. 
}
\label{fig:MXeta}
\vskip .5 cm
\end{figure}

We can impose here partial unification conditions, i.e.~$\alpha_2(M_{X})=\alpha_3(M_{X})$ from Eqs.~(\ref{eq:alphas_DDG}) and get the solution, as a function of $M_c$,
%
\beq
M_{X}=\gamma\, \mathcal W\left[\frac{1}{\gamma} e^{\left( 1+\frac{\pi}{\tilde b_{23}}\alpha^{-1}_{23}(M_c) \right)/\gamma}\right]  M_c\,,\quad\textrm{with}\quad \gamma=\frac{b_{23}-\tilde b_{23}}{2\tilde b_{23}} \,\,,
\label{eq:MX}
\eeq
where we are using the notation $\alpha_{ij}^{-1}(\mu)=\alpha_i^{-1}(\mu)-\alpha_j^{-1}(\mu)$, $b_{ij}=b_i-b_j$, $\tilde b_{ij}=\tilde b_i-\tilde b_j$, and $\mathcal W$ is the Lambert function. 
Notice that as $\tilde b_{23}=11/3$ does not depend on the number of generations in the bulk $\eta$, the unification scale does not depend on it either. Also notice that the expression (\ref{eq:MX}) is exact and no logarithmic term has been neglected therein.
We plot in Fig.~\ref{fig:MXeta} the partial unification scale $M_{X}$ (left panel) as a function of the compactification scale $M_c$.

Full unification can be implemented by imposing the further condition $\alpha_Y(M_X)=\alpha_3(M_X)$ which translates into the condition on the value of $\eta$ as
%
%
\beq
\eta=\frac{3}{23}\, \frac{12\pi \alpha_{Y3}^{-1}(M_c)-128(M_X/M_c-1)-19 \log(M_X/M_c)}{2(M_X/M_c-1)-\log(M_X/M_c)} \,.
\eeq
In the right panel of Fig.~\ref{fig:MXeta} we plot the number of bulk generations $\eta$ ensuring full unification as a function of the compactification scale $M_c$ (as well as the full unification scale $M_X$).
In all cases, the compactification scale is determined by the SM running of gauge couplings. 
Moreover from the right panel of Fig.~\ref{fig:MXeta} we can see that for values $\eta\simeq 3$, low values of $M_c$ are preferred from unification of gauge couplings, while clearly values $\eta<3$ are disfavored;
the value $\eta=3$ appears for $M_c\simeq 640$ GeV, which strongly suggests an approximate unification for values $M_c$ in the few TeV region.

\begin{figure}[htb!]
\vskip .5cm 
 \begin{center}
  \includegraphics[width=7.5 cm]{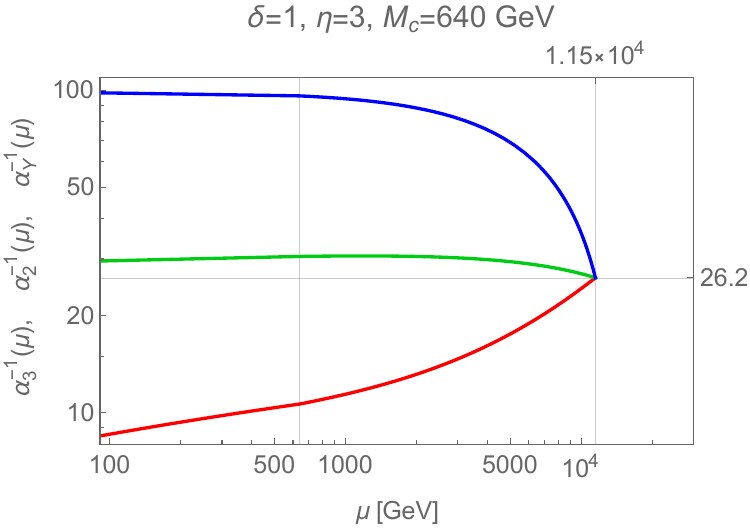}  \,\,\,\,\,\,\,\,\,\,
 \includegraphics[width=7.5 cm]{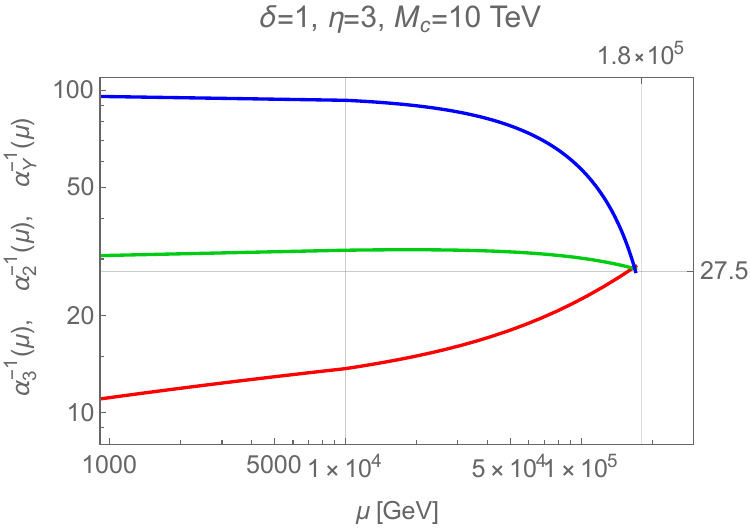}  
 \vskip .5cm
 \end{center}
\caption{\baselineskip=12 pt \small \it 
The case of one extra dimension ($\delta=1$) and three generations in the bulk ($\eta=3$). 
Running of gauge couplings $\alpha_3$, $\alpha_2$ and $\alpha_Y$, for $M_c=640$ GeV (left) and $M_c=10$ TeV (right). 
}
\label{fig:DDG}
\vskip .5 cm
\end{figure}

We plot in Fig.~\ref{fig:DDG} the running of $\alpha_3(\mu)$, $\alpha_2(\mu)$ and $\alpha_Y(\mu)$ for three families of fermions in the bulk, $\eta=3$, and $M_c=640$ GeV (left panel) and $M_c=10$ TeV (right panel).
In the former case, unification is perfect; even in the latter case the gauge couplings unify very well, within a few percent.
The unification happens at $M_X\simeq 10$ TeV for $M_c=640$ GeV, and $M_X\simeq 180$ TeV for $M_c=10$ TeV, consistently with the left panel of Fig.\,\ref{fig:MXeta}.

\begin{figure}[htb!]
\vskip .5cm 
 \begin{center}
  \includegraphics[width=7.5 cm]{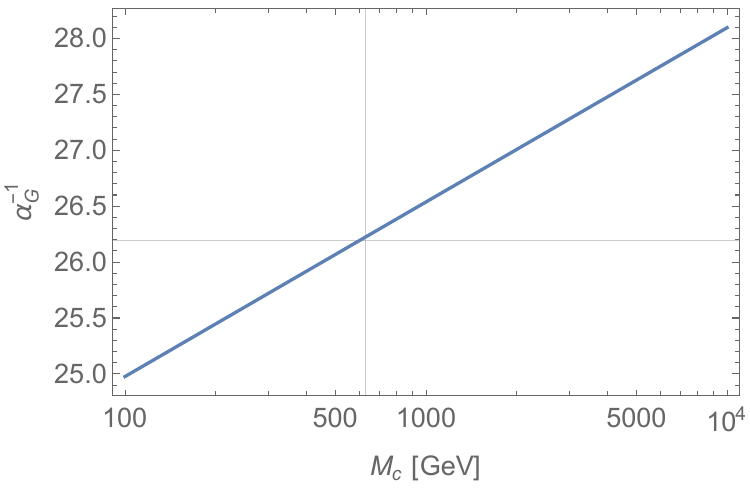}  \,\,\,\,\,
  \includegraphics[width=7.5 cm]{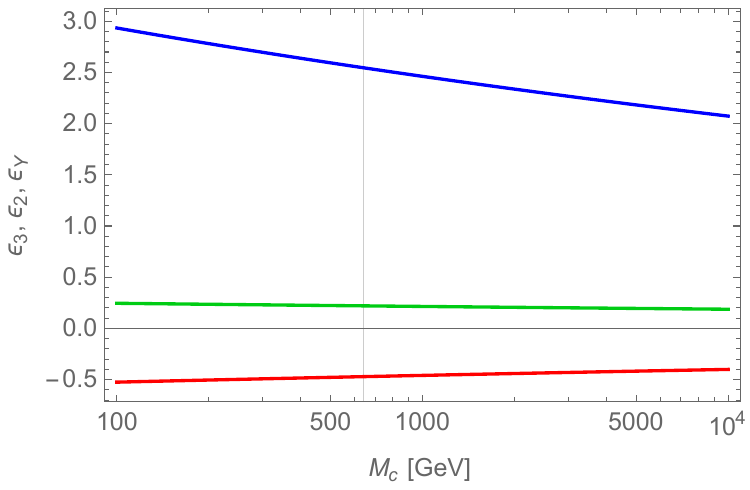}  \,
 \end{center}
\caption{\baselineskip=12 pt \small \it 
The case of one extra dimension ($\delta=1$) and three generations in the bulk ($\eta=3$). The value of $\alpha_G$ (left panel) and the parameters $\epsilon_{3,2,Y}$ (right panel) as functions of $M_c$.
}
\label{fig:epsilonDDG}
\vskip .5 cm
\end{figure}

Notice that unification happens here as the condition
\beq
\alpha_G= \alpha_Y(M_X)=\alpha_2(M_X)=\alpha_3(M_X) \sim 1/27
\label{eq-unif-DDG}
\eeq
which corresponds to values of the Kac-Moody levels $k_Y=k_2=k_3=1$, in contrast to the embedding into a unification group for which $k_Y=5/3$, suggesting a stringy unification origin. 
A plot of $\alpha_{\rm GUT}$ as a function of $M_c$ is shown in the left plot of Fig.~\ref{fig:epsilonDDG}.

In terms of the general parametrization proposed here, the unification condition of Eq.\,(\ref{eq-unif-DDG}), becomes
\beq
\alpha_G= (1+\epsilon_Y) \alpha_Y^{\rm SM}(M_X)= (1+\epsilon_2) \alpha^{\rm SM}_2(M_X)= (1+\epsilon_3) \alpha^{\rm SM}_3(M_X) \,.
\eeq
A plot of the epsilon corrections as functions of $M_c$ is shown in the right panel of Fig.~\ref{fig:epsilonDDG},
and their explicit expressions are given by
\beq
\epsilon_i^{-1}=\frac{\alpha_i^{-1}(M_c)-\frac{b_i}{2\pi}\log(M_X/M_c)}{\frac{\tilde b_i}{\pi}(M_X/M_c-1)-\frac{\tilde b_i}{2\pi} \log(M_X/M_c)}-1
\eeq
where $M_X$ is fixed by Eq.~(\ref{eq:MX}). The contribution proportional to the logarithm is generically subleading. Notice also that $\epsilon_3$ turns out to be negative.

Theoretical and phenomenological issues raised in these theories, as perturbativity, higher-loop corrections, sensitivity to the unification scale thresholds, proton decay and embedding into string theory, have been addressed in the original literature~\cite{Dienes:1998vg}. The main obstacle to string GUT models with a unification simple group $G$, leaving aside the phenomenological problem of proton decay in theories with low scale unification, is that to break $G$ to the Standard Model group one needs a Higgs in the adjoint representation, which is hard to obtain given the implied requirement that $k_G\geq 2$. In fact GUT unification is not required in string theories, as unification is a consequence of the relation $8\pi G_N/\alpha'=g_i^2k_i=g_{\rm string}^2$ ($i=Y,2,3$), where $G_N$ is the Newton constant, $\alpha'$ the string tension, $g_i$ the gauge coupling of group $G_i$ and $g_{\rm string}$ the string coupling.
This leads us to consider as path to unification the case of considering an appropriate affine level $k_Y<5/3$.  The presence in the spectrum of a right-handed lepton with hypercharge $|Y_{\ell_R}|=1$  leads to the condition $k_Y\geq 1$. MSSM-like models with $k_Y<5/3$ have been identified in Ref.~\cite{Dienes:1995sq}. Non-supersymmetric intersecting branes string theories with the Standard Model content have been worked out in Ref.~\cite{Ibanez:2001nd}.

The previous analysis is based on the SM hypercharge normalization $k_Y=1$ with the non-abelian Kac-Moody levels $k_2=k_3=1$. The possibility of getting different values of Kac-Moody levels, in particular for the hypercharge, $k_Y$, for a broad class of string models was achieved in Ref.~\cite{Dienes:1995sq}. The obtained result is that for most realistic string models it turns out that $k_Y\geq 5/3$, while particular values $k_Y<5/3$ can be obtained in the presence of singularities (e.g.~in orbifold compactifications) and/or for higher non-abelian Kac-Moody levels. While a detailed analysis in terms of string theories is beyond the scope of the present paper, we have made an analysis of possible values of $k_Y$ and $M_c$ which predict unification depending on the different number of fermions generation propagating in the bulk.

\begin{figure}[htb!]
\vskip .5cm 
 \begin{center}
  \includegraphics[width=5.1 cm]{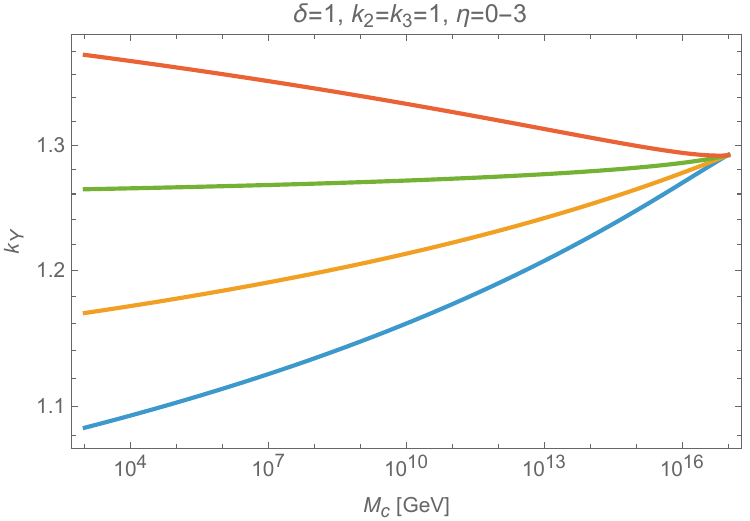} 
  \includegraphics[width=5.1 cm]{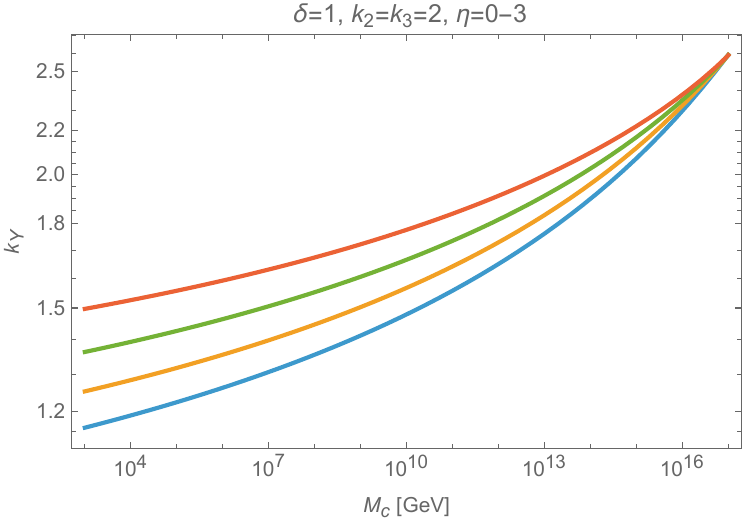}  \,
 \includegraphics[width=5.1 cm]{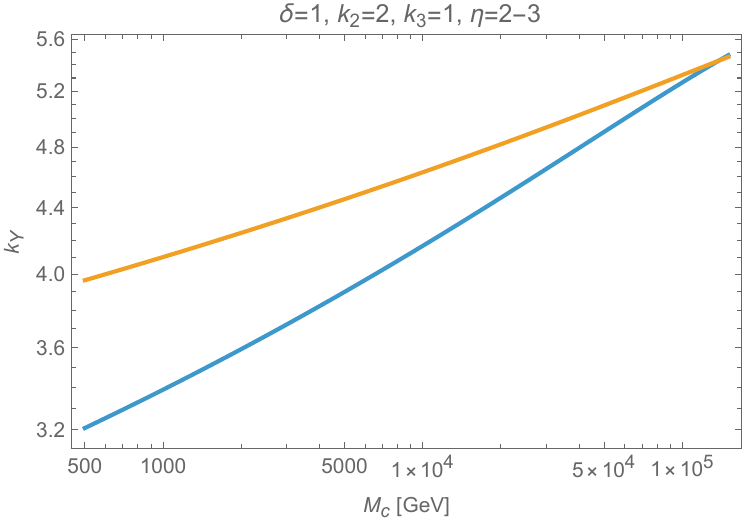}   \vskip .5cm
 \end{center}
\caption{\baselineskip=12 pt \small \it The prediction of $k_Y$ as a function of $M_c$ from gauge coupling unification for the cases with $k_2=k_3=1$ (left panel), $k_2=k_3=2$ (middle panel) and $k_2=2$, $k_3=1$ (right panels). In every panel the upper line corresponds to $\eta=0$, and the lower line to $\eta=3$. 
}
\label{fig:kY}
\vskip .5 cm
\end{figure}

The results are shown in the plots of Fig.~\ref{fig:kY} for the cases with $k_2=k_3=1$ (left panel), $k_2=k_3=2$ (middle panel) and $k_2=2$, $k_3=1$ (right panel). We have presented the predictions for no fermion generation in the bulk $\eta=0$ (upper lines) and a number of fermion generations $\eta\leq 3$ (subsequent lines); in the last case, for $\eta=0,1$ there is no solution to unification for any value of $k_Y$. 
As we can see from the left plot, for the case of basic non-abelian Kac-Moody levels, unification can only be achieved for $k_Y<5/3$. However, for higher non-abelian Kac-Moody levels unification can also be achieved for  $k_Y>5/3$. In the first two cases there is always a point in the plane $(M_c,k_Y)$ for which unification holds independently of the number of fermion generations propagating in the bulk. In the third case there is also a point where unification holds independently of the number of bulk generation, although here only the values $\eta=2,3$ are consistent with unification.
For all the plots, what happens at such point is that $M_X=M_c$ (so larger values of $M_c$ for which $M_X<M_c$, are forbidden):
the unification point is when the 4D SM (with the particular values of $k_2$ and $k_3$) unifies by itself. For $k_2=k_3=1$ the unification point happens at $M_X=\mu_{32}^{(1)}$
which yields the well known relation $k_Y=1.28$. For the case $k_2=k_3=2$ the unification point is also at $M_X=\mu_{32}^{(1)}$ and $k_Y=2.6$. Finally for the case $k_2=2, k_3=1$ the unification happens for $M_X\simeq 10^5$ GeV, in agreement with the result from the right panel of Fig.~\ref{fig-k2k3}, and $k_Y=5.4$.

\section{Conclusions}
\label{sec-concl}

In this work we have explored the role of the SM non-Abelian gauge coupling partial unification scale, $\mu^{\rm SM}_{32}\approx 2.8 \times 10^{16}$ GeV, 
in relation to the issue of full or partial unification in models of new physics, with or without a desert below the unification scale.  

Starting from the fact that $\mu^{\rm SM}_{32}$ is interestingly close to the (nearly full) GCU predicted in the low energy SUSY framework, $M^{\rm SUSY}_X \sim 2 \times 10^{16}$ GeV,
we investigated the associated beta functions. We noticed that, similarly, also the 2HDM and the split-SUSY models predict the partial unification of their non-Abelian gauge 
couplings at (or close to) $\mu^{\rm SM}_{32}$. 

We then introduced a simple general parametrization, Eq.\,(\ref{eq-eps-gen}), useful to understand how (full or partial) GCU happens in any new physics model. 
In particular, we emphasized that the corrections to the non-Abelian gauge couplings, denoted by $\epsilon_2$ and $\epsilon_3$, 
control the closeness of the unification scale $M_X$ to $\mu^{\rm SM}_{32}$.

If the non-Abelian corrections are equal (or very close), then $M_X$ is equal (or very close) to $\mu^{\rm SM}_{32}$, as happens for low energy SUSY and other relevant models, including 
the 2HDM and split-SUSY models. For this reason, all models of this type will be denoted as mirage SUSY; 
the value of the (partially or fully) unified gauge coupling can however sizably differ with respect to low energy SUSY, as well as among the mirage SUSY models.

Only in the case that the non-Abelian corrections are sizably different, can one disentangle $M_X$ from $\mu^{\rm SM}_{32}$,
as the quantity $\log (M_X/\mu_{32}^{\rm SM})$ turns out to be proportional to the difference $\epsilon_3-\epsilon_2$, see Eq.\,(\ref{eq-MX}) and Fig.\,\ref{fig-MX}.

Models of the non-desert type usually have positive and similar non-Abelian corrections, $\epsilon_3 \approx \epsilon_2$; 
this corresponds to the dot-dashed line in the plot of Fig.\,\ref{fig-MX}.
Many relevant models lie close to this line, including low energy SUSY, 2HDM and split-SUSY.  
They thus belong to the category of mirage SUSY models, for which the partial (or full) GCU scale is close to $\mu_{32}^{\rm SM}$.

Models of the desert type have corrections stemming from UV effects. Here we postponed the interesting (but lengthy) case of effective corrections from a non-renormalizable kinetic term,
and rather considered in some detail the case in which the corrections satisfy string-inspired relations.
For the latter scenario, we found that, in addition to the possibility of full GCU at $\mu_{32}^{\rm SM}$ (obtained when $\epsilon_3=\epsilon_2=0$), 
there is also the possibility of a much lower scale, $M_X\approx 100$\,TeV, obtained when $\epsilon_3=0$ and $\epsilon_2=1$.
The latter scale is remarkably close to the energy range accessible to future experiments. 
The phenomenology of the latter scenario, which corresponds to the upper left corner in Fig. \ref{fig-MX}, deserves further consideration. 
In addition, we focused on models with power-low running~\cite{Dienes:1998vh,Dienes:1998vg,Dienes:1998qh}, outlining an interesting connection between coupling unification and the number of fermion families propagating in the bulk of the extra dimensions.

As the non-Abelian gauge groups in the SM unify at a pretty high scale $\mu^{\rm SM}_{32}$, one could be led to the appealing assumption that new physics deals mainly with extra hypercharge (weakly coupled) fields. However, we have shown that this case requires $\epsilon_1<0$, which is not compatible with the presence of extra fields with masses below the unification scale. In fact, this case would only be consistent with threshold effects from supermassive states of some UV completion of the theory, 
as e.g.~the string excitations for the case where the string scale is $M_s\gtrsim \mu_{32}^{\rm SM}$, or Kaluza-Klein modes corresponding to dimensions with inverse size $1/R\gtrsim \mu_{32}^{\rm SM}$. The possibility of modifying the kinetic terms of the gauge bosons by higher dimensional operators will be analyzed elsewhere.

Summarizing, we have shown that the SM partial unification scale $\mu^{\rm SM}_{32}$ is a useful guide for new physics models displaying full or partial gauge coupling unification.

\vskip 1.cm

\section*{\large Acknowledgments}

We are indebted to Keith Dienes for pointing out the possibility of power-law unification, missing from the initial version of the paper.
We thank the CERN Theory Department for kind hospitality and support during the completion of this work. 
IM acknowledges partial support by the research project TAsP (Theoretical Astroparticle Physics) funded by the Istituto Nazionale di Fisica Nucleare (INFN). 
The work of MQ is supported in part by the R\&D\&i project PID2023-146686NB-C31, funded by MICIU/AEI/10.13039/501100011033/ and by ERDF/EU. IFAE is partially funded by the CERCA program of the Generalitat de Catalunya.

\appendix
\vskip 1.cm

\appendix
\numberwithin{equation}{section}

\bibliographystyle{elsarticle-num} 
\bibliography{bib-SMunif} 

@article{Dienes:1995sq,
    author = "Dienes, Keith R. and Faraggi, Alon E. and March-Russell, John",
    title = "{String unification, higher level gauge symmetries, and exotic hypercharge normalizations}",
    eprint = "hep-th/9510223",
    archivePrefix = "arXiv",
    reportNumber = "IASSNS-HEP-95-25, UFIFT-HEP-95-26, NSF-ITP-95-130",
    doi = "10.1016/0550-3213(96)00085-5",
    journal = "Nucl. Phys. B",
    volume = "467",
    pages = "44--99",
    year = "1996"
}

@article{Ibanez:2001nd,
    author = "Ibanez, Luis E. and Marchesano, F. and Rabadan, R.",
    title = "{Getting just the standard model at intersecting branes}",
    eprint = "hep-th/0105155",
    archivePrefix = "arXiv",
    reportNumber = "FTUAM-01-09, IFT-UAM-CSIC-01-15",
    doi = "10.1088/1126-6708/2001/11/002",
    journal = "JHEP",
    volume = "11",
    pages = "002",
    year = "2001"
}

@article{Delgado:1999ba,
    author = "Delgado, A. and Quiros, M.",
    title = "{Strong coupling unification and extra dimensions}",
    eprint = "hep-ph/9903400",
    archivePrefix = "arXiv",
    reportNumber = "IEM-FT-189-99, IFT-UAM-CSIC-99-9",
    doi = "10.1016/S0550-3213(99)00440-X",
    journal = "Nucl. Phys. B",
    volume = "559",
    pages = "235--254",
    year = "1999"
}

@inproceedings{Dienes:1998qh,
    author = "Dienes, Keith R. and Dudas, Emilian and Gherghetta, Tony",
    title = "{TeV scale GUTs}",
    booktitle = "{1st European Meeting From the Planck Scale to the Electroweak Scale}",
    eprint = "hep-ph/9807522",
    archivePrefix = "arXiv",
    reportNumber = "CERN-TH-98-245",
    pages = "613--620",
    month = "7",
    year = "1998"
}

@article{Dienes:1998vg,
    author = "Dienes, Keith R. and Dudas, Emilian and Gherghetta, Tony",
    title = "{Grand unification at intermediate mass scales through extra dimensions}",
    eprint = "hep-ph/9806292",
    archivePrefix = "arXiv",
    reportNumber = "CERN-TH-98-100",
    doi = "10.1016/S0550-3213(98)00669-5",
    journal = "Nucl. Phys. B",
    volume = "537",
    pages = "47--108",
    year = "1999"
}

@article{Dienes:1998vh,
    author = "Dienes, Keith R. and Dudas, Emilian and Gherghetta, Tony",
    title = "{Extra space-time dimensions and unification}",
    eprint = "hep-ph/9803466",
    archivePrefix = "arXiv",
    reportNumber = "CERN-TH-98-065",
    doi = "10.1016/S0370-2693(98)00977-0",
    journal = "Phys. Lett. B",
    volume = "436",
    pages = "55--65",
    year = "1998"
}

@article{Helsens:2019bfw,
    author = "Helsens, Clement and Jamin, David and Mangano, Michelangelo L. and Rizzo, Thomas G. and Selvaggi, Michele",
    title = "{Heavy resonances at energy-frontier hadron colliders}",
    eprint = "1902.11217",
    archivePrefix = "arXiv",
    primaryClass = "hep-ph",
    reportNumber = "CERN-TH-2019-020, SLAC-PUB-17408",
    doi = "10.1140/epjc/s10052-019-7062-3",
    journal = "Eur. Phys. J. C",
    volume = "79",
    pages = "569",
    year = "2019"
}

@article{Antoniadis:1998ig,
    author = "Antoniadis, Ignatios and Arkani-Hamed, Nima and Dimopoulos, Savas and Dvali, G. R.",
    title = "{New dimensions at a millimeter to a Fermi and superstrings at a TeV}",
    eprint = "hep-ph/9804398",
    archivePrefix = "arXiv",
    reportNumber = "SLAC-PUB-7801, SU-ITP-98-28, CPTH-S608-0498, IC-98-39",
    doi = "10.1016/S0370-2693(98)00860-0",
    journal = "Phys. Lett. B",
    volume = "436",
    pages = "257--263",
    year = "1998"
}

@article{Arkani-Hamed:1998jmv,
    author = "Arkani-Hamed, Nima and Dimopoulos, Savas and Dvali, G. R.",
    title = "{The Hierarchy problem and new dimensions at a millimeter}",
    eprint = "hep-ph/9803315",
    archivePrefix = "arXiv",
    reportNumber = "SLAC-PUB-7769, SU-ITP-98-13",
    doi = "10.1016/S0370-2693(98)00466-3",
    journal = "Phys. Lett. B",
    volume = "429",
    pages = "263--272",
    year = "1998"
}

@article{Antoniadis:2011qw,
    author = "Antoniadis, Ignatios and Arvanitaki, Asimina and Dimopoulos, Savas and Giveon, Amit",
    title = "{Phenomenology of TeV Little String Theory from Holography}",
    eprint = "1102.4043",
    archivePrefix = "arXiv",
    primaryClass = "hep-ph",
    reportNumber = "CERN-PH-TH-2011-024",
    doi = "10.1103/PhysRevLett.108.081602",
    journal = "Phys. Rev. Lett.",
    volume = "108",
    pages = "081602",
    year = "2012"
}

@article{Antoniadis:2001sw,
    author = "Antoniadis, Ignatios and Dimopoulos, Savas and Giveon, Amit",
    title = "{Little string theory at a TeV}",
    eprint = "hep-th/0103033",
    archivePrefix = "arXiv",
    reportNumber = "CERN-TH-2001-066, RI-07-00, ITP-01-04",
    doi = "10.1088/1126-6708/2001/05/055",
    journal = "JHEP",
    volume = "05",
    pages = "055",
    year = "2001"
}

@article{Kutasov:2001uf,
    author = "Kutasov, D.",
    editor = "Bachas, C. and Maldacena, Juan Martin and Narain, K. S. and Randjbar-Daemi, S.",
    title = "{Introduction to little string theory}",
    journal = "ICTP Lect. Notes Ser.",
    volume = "7",
    pages = "165--209",
    year = "2002"
}

@article{Dimopoulos:1981yj,
    author = "Dimopoulos, S. and Raby, S. and Wilczek, Frank",
    title = "{Supersymmetry and the Scale of Unification}",
    reportNumber = "NSF-ITP-81-31",
    doi = "10.1103/PhysRevD.24.1681",
    journal = "Phys. Rev. D",
    volume = "24",
    pages = "1681--1683",
    year = "1981"
}

@article{Dimopoulos:1981zb,
    author = "Dimopoulos, Savas and Georgi, Howard",
    title = "{Softly Broken Supersymmetry and SU(5)}",
    reportNumber = "HUTP-81/A022",
    doi = "10.1016/0550-3213(81)90522-8",
    journal = "Nucl. Phys. B",
    volume = "193",
    pages = "150--162",
    year = "1981"
}

@article{Ibanez:1981yh,
    author = "Ibanez, Luis E. and Ross, Graham G.",
    title = "{Low-Energy Predictions in Supersymmetric Grand Unified Theories}",
    reportNumber = "OXFORD-TP 65/81",
    doi = "10.1016/0370-2693(81)91200-4",
    journal = "Phys. Lett. B",
    volume = "105",
    pages = "439--442",
    year = "1981"
}

@article{Sakai:1981gr,
    author = "Sakai, N.",
    title = "{Naturalness in Supersymmetric Guts}",
    reportNumber = "TU/81/225",
    doi = "10.1007/BF01573998",
    journal = "Z. Phys. C",
    volume = "11",
    pages = "153",
    year = "1981"
}

@article{Einhorn:1981sx,
    author = "Einhorn, M. B. and Jones, D. R. T.",
    title = "{The Weak Mixing Angle and Unification Mass in Supersymmetric SU(5)}",
    reportNumber = "UM HE 81-55",
    doi = "10.1016/0550-3213(82)90502-8",
    journal = "Nucl. Phys. B",
    volume = "196",
    pages = "475--488",
    year = "1982"
}

@article{Marciano:1981un,
    author = "Marciano, William J. and Senjanovic, Goran",
    title = "{Predictions of Supersymmetric Grand Unified Theories}",
    reportNumber = "Print-81-0844 (BROOKHAVEN), BNL-30398",
    doi = "10.1103/PhysRevD.25.3092",
    journal = "Phys. Rev. D",
    volume = "25",
    pages = "3092",
    year = "1982"
}

@article{Amaldi:1991cn,
    author = "Amaldi, Ugo and de Boer, Wim and Furstenau, Hermann",
    title = "{Comparison of grand unified theories with electroweak and strong coupling constants measured at LEP}",
    reportNumber = "CERN-PPE-91-44, IEKP-KA-91-01",
    doi = "10.1016/0370-2693(91)91641-8",
    journal = "Phys. Lett. B",
    volume = "260",
    pages = "447--455",
    year = "1991"
}

@article{Ellis:1990wk,
    author = "Ellis, John R. and Kelley, S. and Nanopoulos, Dimitri V.",
    title = "{Probing the desert using gauge coupling unification}",
    reportNumber = "CERN-TH-5943-90, CTP-TAMU-97-90, ACT-19",
    doi = "10.1016/0370-2693(91)90980-5",
    journal = "Phys. Lett. B",
    volume = "260",
    pages = "131--137",
    year = "1991"
}

@article{Langacker:1991an,
    author = "Langacker, Paul and Luo, Ming-xing",
    title = "{Implications of precision electroweak experiments for $M_t$, $\rho_{0}$, $\sin^2\theta_W$ and grand unification}",
    reportNumber = "UPR-0466T",
    doi = "10.1103/PhysRevD.44.817",
    journal = "Phys. Rev. D",
    volume = "44",
    pages = "817--822",
    year = "1991"
}

@article{Giunti:1991ta,
    author = "Giunti, C. and Kim, C. W. and Lee, U. W.",
    title = "{Running coupling constants and grand unification models}",
    doi = "10.1142/S0217732391001883",
    journal = "Mod. Phys. Lett. A",
    volume = "6",
    pages = "1745--1755",
    year = "1991"
}

@article{ParticleDataGroup:2024cfk,
    author = "Navas, S. and others",
    collaboration = "Particle Data Group",
    title = "{Review of particle physics}",
    doi = "10.1103/PhysRevD.110.030001",
    journal = "Phys. Rev. D",
    volume = "110",
    number = "3",
    pages = "030001",
    year = "2024"
}

@article{Senjanovic:2023jvv,
    author = "Senjanovi\'c, Goran and Zantedeschi, Michael",
    title = "{Grand unification, small vs large representations, hadron colliders and all that}",
    eprint = "2304.07932",
    archivePrefix = "arXiv",
    primaryClass = "hep-ph",
    doi = "10.22323/1.436.0127",
    journal = "PoS",
    volume = "CORFU2022",
    pages = "127",
    year = "2023"
}

@article{Arkani-Hamed:2004ymt,
    author = "Arkani-Hamed, Nima and Dimopoulos, Savas",
    title = "{Supersymmetric unification without low energy supersymmetry and signatures for fine-tuning at the LHC}",
    eprint = "hep-th/0405159",
    archivePrefix = "arXiv",
    doi = "10.1088/1126-6708/2005/06/073",
    journal = "JHEP",
    volume = "06",
    pages = "073",
    year = "2005"
}

@article{Giudice:2004tc,
    author = "Giudice, G. F. and Romanino, A.",
    title = "{Split supersymmetry}",
    eprint = "hep-ph/0406088",
    archivePrefix = "arXiv",
    reportNumber = "CERN-PH-TH-2004-100",
    doi = "10.1016/j.nuclphysb.2004.08.001",
    journal = "Nucl. Phys. B",
    volume = "699",
    pages = "65--89",
    year = "2004",
    note = "[Erratum: Nucl.Phys.B 706, 487--487 (2005)]"
}

@article{Lee:1973iz,
    author = "Lee, T. D.",
    editor = "Feinberg, G.",
    title = "{A Theory of Spontaneous T Violation}",
    doi = "10.1103/PhysRevD.8.1226",
    journal = "Phys. Rev. D",
    volume = "8",
    pages = "1226--1239",
    year = "1973"
}

@article{Branco:2011iw,
    author = "Branco, G. C. and Ferreira, P. M. and Lavoura, L. and Rebelo, M. N. and Sher, Marc and Silva, Joao P.",
    title = "{Theory and phenomenology of two-Higgs-doublet models}",
    eprint = "1106.0034",
    archivePrefix = "arXiv",
    primaryClass = "hep-ph",
    doi = "10.1016/j.physrep.2012.02.002",
    journal = "Phys. Rept.",
    volume = "516",
    pages = "1--102",
    year = "2012"
}

@article{Langacker:1992rq,
    author = "Langacker, Paul and Polonsky, Nir",
    title = "{Uncertainties in coupling constant unification}",
    eprint = "hep-ph/9210235",
    archivePrefix = "arXiv",
    reportNumber = "UPR-0513-T, UPR-0513T",
    doi = "10.1103/PhysRevD.47.4028",
    journal = "Phys. Rev. D",
    volume = "47",
    pages = "4028--4045",
    year = "1993"
}

@article{Carena:1993ag,
    author = "Carena, Marcela and Pokorski, S. and Wagner, C. E. M.",
    title = "{On the unification of couplings in the minimal supersymmetric Standard Model}",
    eprint = "hep-ph/9303202",
    archivePrefix = "arXiv",
    reportNumber = "MPI-PH-93-10",
    doi = "10.1016/0550-3213(93)90161-H",
    journal = "Nucl. Phys. B",
    volume = "406",
    pages = "59--89",
    year = "1993"
}

@article{Altarelli:2000fu,
    author = "Altarelli, Guido and Feruglio, Ferruccio and Masina, Isabella",
    title = "{From minimal to realistic supersymmetric SU(5) grand unification}",
    eprint = "hep-ph/0007254",
    archivePrefix = "arXiv",
    reportNumber = "CERN-TH-2000-171, DFPD-00-TH-34",
    doi = "10.1088/1126-6708/2000/11/040",
    journal = "JHEP",
    volume = "11",
    pages = "040",
    year = "2000"
}

@article{Masina:2001pp,
    author = "Masina, Isabella",
    title = "{The Problem of neutrino masses in extensions of the standard model}",
    eprint = "hep-ph/0107220",
    archivePrefix = "arXiv",
    reportNumber = "DFPD-01-TH-30",
    doi = "10.1142/S0217751X01005456",
    journal = "Int. J. Mod. Phys. A",
    volume = "16",
    pages = "5101--5200",
    year = "2001"
}

@article{Kehagias:2005vz,
    author = "Kehagias, Alex and Tracas, N. D.",
    title = "{Standard and non-standard extra matter for non-supersymmetric unification}",
    eprint = "hep-ph/0506144",
    archivePrefix = "arXiv",
    month = "6",
    year = "2005"
}

@article{DiLuzio:2013dda,
    author = "Di Luzio, Luca and Mihaila, Luminita",
    title = "{Unification scale vs. electroweak-triplet mass in the $SU(5) + 24_F$ model at three loops}",
    eprint = "1305.2850",
    archivePrefix = "arXiv",
    primaryClass = "hep-ph",
    doi = "10.1103/PhysRevD.87.115025",
    journal = "Phys. Rev. D",
    volume = "87",
    pages = "115025",
    year = "2013"
}

@article{Ellis:2015jwa,
    author = "Ellis, Sebastian A. R. and Wells, James D.",
    title = "{Visualizing gauge unification with high-scale thresholds}",
    eprint = "1502.01362",
    archivePrefix = "arXiv",
    primaryClass = "hep-ph",
    doi = "10.1103/PhysRevD.91.075016",
    journal = "Phys. Rev. D",
    volume = "91",
    number = "7",
    pages = "075016",
    year = "2015"
}

@article{Schwichtenberg:2018cka,
    author = "Schwichtenberg, Jakob",
    title = "{Gauge Coupling Unification without Supersymmetry}",
    eprint = "1808.10329",
    archivePrefix = "arXiv",
    primaryClass = "hep-ph",
    reportNumber = "TTP18-031",
    doi = "10.1140/epjc/s10052-019-6878-1",
    journal = "Eur. Phys. J. C",
    volume = "79",
    number = "4",
    pages = "351",
    year = "2019"
}

@article{Meloni:2019jcf,
    author = "Meloni, Davide and Ohlsson, Tommy and Pernow, Marcus",
    title = "{Threshold effects in SO(10) models with one intermediate breaking scale}",
    eprint = "1911.11411",
    archivePrefix = "arXiv",
    primaryClass = "hep-ph",
    doi = "10.1140/epjc/s10052-020-8308-9",
    journal = "Eur. Phys. J. C",
    volume = "80",
    number = "9",
    pages = "840",
    year = "2020"
}

@article{Djouadi:2022gws,
    author = "Djouadi, Abdelhak and Fonseca, Renato and Ouyang, Ruiwen and Raidal, Martti",
    title = "{Non-supersymmetric SO(10) models with Gauge and Yukawa coupling unification}",
    eprint = "2212.11315",
    archivePrefix = "arXiv",
    primaryClass = "hep-ph",
    doi = "10.1140/epjc/s10052-023-11696-4",
    journal = "Eur. Phys. J. C",
    volume = "83",
    number = "6",
    pages = "529",
    year = "2023"
}

@article{Haba:2024lox,
    author = "Haba, Naoyuki and Nagano, Keisuke and Shimizu, Yasuhiro and Yamada, Toshifumi",
    title = "{Gauge Coupling Unification and Proton Decay via 45 Higgs Boson in SU(5) GUT}",
    eprint = "2402.15124",
    archivePrefix = "arXiv",
    primaryClass = "hep-ph",
    doi = "10.1093/ptep/ptae066",
    journal = "PTEP",
    volume = "2024",
    number = "5",
    pages = "053B05",
    year = "2024"
}

@article{Masina:2024ybn,
    author = "Masina, Isabella and Quiros, Mariano",
    title = "{Electroweak metastability and Higgs inflation}",
    eprint = "2403.02461",
    archivePrefix = "arXiv",
    primaryClass = "hep-ph",
    doi = "10.1140/epjc/s10052-024-13522-x",
    journal = "Eur. Phys. J. C",
    volume = "84",
    number = "11",
    pages = "1153",
    year = "2024"
}

@article{Alam:2022cdv,
    author = "Alam, Zamiul and Martin, Stephen P.",
    title = "{Standard model at 200~GeV}",
    eprint = "2211.08576",
    archivePrefix = "arXiv",
    primaryClass = "hep-ph",
    doi = "10.1103/PhysRevD.107.013010",
    journal = "Phys. Rev. D",
    volume = "107",
    number = "1",
    pages = "013010",
    year = "2023"
}

@article{Dienes:1996du,
    author = "Dienes, Keith R.",
    title = "{String theory and the path to unification: A Review of recent developments}",
    eprint = "hep-th/9602045",
    archivePrefix = "arXiv",
    reportNumber = "IASSNS-HEP-95-97, IASSNS-HEP-95-97-(SEPTEMBER-1996)",
    doi = "10.1016/S0370-1573(97)00009-4",
    journal = "Phys. Rept.",
    volume = "287",
    pages = "447--525",
    year = "1997"
}

@article{Cho:1997gm,
    author = "Cho, Gi-Chol and Hagiwara, Kaoru",
    title = "{String unification scale and the hypercharge Kac-Moody level in the nonsupersymmetric standard model}",
    eprint = "hep-ph/9709279",
    archivePrefix = "arXiv",
    reportNumber = "KEK-TH-537, KEK-PREPRINT-97-144",
    doi = "10.1016/S0370-2693(97)01488-3",
    journal = "Phys. Lett. B",
    volume = "419",
    pages = "199--205",
    year = "1998"
}

@article{Heckman:2024trz,
    author = "Heckman, Jonathan J. and Vafa, Cumrun and Weigand, Timo and Xu, Fengjun",
    title = "{Dark dimension and the grand unification of forces}",
    eprint = "2409.01405",
    archivePrefix = "arXiv",
    primaryClass = "hep-th",
    reportNumber = "ZMP-HH/24-19",
    doi = "10.1103/PhysRevD.111.046014",
    journal = "Phys. Rev. D",
    volume = "111",
    number = "4",
    pages = "046014",
    year = "2025"
}

@article{Anchordoqui:2025nmb,
    author = "Anchordoqui, Luis A. and Antoniadis, Ignatios and Lust, Dieter",
    title = "{Two Micron-Size Dark Dimensions}",
    eprint = "2501.11690",
    archivePrefix = "arXiv",
    primaryClass = "hep-th",
    reportNumber = "MPP-2025-5, LMU-ASC 02/25",
    doi = "10.1002/prop.70015",
    journal = "Fortsch. Phys.",
    volume = "73",
    number = "8",
    pages = "e70015",
    year = "2025"
}

@article{Antoniadis:2025pet,
    author = "Antoniadis, Ignatios and Chatrabhuti, Auttakit and Cunat, Jules and Isono, Hiroshi",
    title = "{Bispectrum from five-dimensional inflation}",
    eprint = "2505.14225",
    archivePrefix = "arXiv",
    primaryClass = "hep-ph",
    doi = "10.1007/JHEP08(2025)163",
    journal = "JHEP",
    volume = "08",
    pages = "163",
    year = "2025"
}

@article{Ettengruber:2025kzw,
    author = "Ettengruber, Manuel and Kuhnel, Florian",
    title = "{Micro Black Hole Dark Matter}",
    eprint = "2506.14871",
    archivePrefix = "arXiv",
    primaryClass = "hep-th",
    month = "6",
    year = "2025"
}

@article{Antoniadis:2025rck,
    author = "Antoniadis, Ignatios and Chatrabhuti, Auttakit and Isono, Hiroshi",
    title = "{Searching for a Dark Dimension Right-handed Neutrino in KATRIN}",
    eprint = "2509.05233",
    archivePrefix = "arXiv",
    primaryClass = "hep-ph",
    month = "9",
    year = "2025"
}

@article{Hill:1983xh,
    author = "Hill, Christopher T.",
    title = "{Are There Significant Gravitational Corrections to the Unification Scale?}",
    reportNumber = "FERMILAB-PUB-83-078-THY",
    doi = "10.1016/0370-2693(84)90451-9",
    journal = "Phys. Lett. B",
    volume = "135",
    pages = "47--51",
    year = "1984"
}

@article{Shafi:1983gz,
    author = "Shafi, Q. and Wetterich, C.",
    title = "{Modification of {GUT} Predictions in the Presence of Spontaneous Compactification}",
    reportNumber = "BA-83-43",
    doi = "10.1103/PhysRevLett.52.875",
    journal = "Phys. Rev. Lett.",
    volume = "52",
    pages = "875",
    year = "1984"
}

@article{Panagiotakopoulos:1984wf,
    author = "Panagiotakopoulos, C. and Shafi, Q.",
    title = "{Dimension Five Interactions, Fermion Masses and Higgs Mediated Proton Decay in SU(5)}",
    reportNumber = "BA-84-13",
    doi = "10.1103/PhysRevLett.52.2336",
    journal = "Phys. Rev. Lett.",
    volume = "52",
    pages = "2336",
    year = "1984"
}

@article{Hall:1992kq,
    author = "Hall, Lawrence J. and Sarid, Uri",
    title = "{Gravitational smearing of minimal supersymmetric unification predictions}",
    eprint = "hep-ph/9210240",
    archivePrefix = "arXiv",
    reportNumber = "LBL-32905, UCB-PTH-92-37",
    doi = "10.1103/PhysRevLett.70.2673",
    journal = "Phys. Rev. Lett.",
    volume = "70",
    pages = "2673--2676",
    year = "1993"
}
\end{document}